\documentclass{aastex631}

\newcommand{\add}[1]{{#1}}
\newcommand{\rev}[1]{{#1}}

\shorttitle{Cosmic Ray Spectral Variations}
\shortauthors{Muangha et al.}

\begin{document}

\title{Variations in the \rev{Inferred Cosmic-Ray Spectral Index} as Measured by Neutron Monitors in Antarctica}

\author{Pradiphat Muangha}
\affiliation{Department of Physics, Faculty of Science, Mahidol University, Bangkok 10400, Thailand}

\author{David Ruffolo}
\affiliation{Department of Physics, Faculty of Science, Mahidol University, Bangkok 10400, Thailand}

\author{Alejandro S\'aiz}
\affiliation{Department of Physics, Faculty of Science, Mahidol University, Bangkok 10400, Thailand}

\author{Chanoknan Banglieng}
\affiliation{Division of Physics, Faculty of Science and Technology, Rajamangala University of Technology Thanyaburi, \\ Pathum Thani 12110, Thailand }

\author{Paul Evenson}
\affiliation{Department of Physics and Astronomy, University of Delaware, Newark, DE 19716, USA}

\author{Surujhdeo Seunarine}
\affiliation{Department of Physics, University of Wisconsin
River Falls, River Falls, WI 54022, USA}

\author{Suyeon Oh}
\affiliation{Department of Earth Science Education, Chonnam National University, Gwangju 61186, South Korea}

\author{Jongil Jung}
\affiliation{Korea Astronomy and Space Science Institute, Daejeon 34055, South Korea}

\author{Marc Duldig}
\affiliation{School of Natural Sciences, University of Tasmania, Hobart, Tasmania 7001, Australia}

\author{John Humble}
\affiliation{School of Natural Sciences, University of Tasmania, Hobart, Tasmania 7001, Australia}

\accepted{by the Astrophysical Journal, 2024 Aug 25}

 \begin{abstract}
A technique has recently been developed for tracking short-term spectral variations in Galactic cosmic rays (GCRs) using data from a single neutron monitor (NM), by collecting histograms of the time delay between successive neutron counts and extracting the leader fraction $L$ as a proxy of the spectral index.
Here we analyze $L$ from four Antarctic NMs during 2015 March to 2023 September.
We have calibrated $L$ from the South Pole NM with respect to a daily spectral index determined from published data of GCR proton fluxes during 2015--2019 from the Alpha Magnetic Spectrometer (AMS-02) aboard the International Space Station. 
Our results demonstrate a robust correlation between the leader fraction and the spectral index fit over the rigidity range 2.97--16.6 GV for AMS-02 data, with uncertainty 0.018 in the daily spectral index as inferred from $L$. 
In addition to the 11-year solar activity cycle, a wavelet analysis confirms a 27-day periodicity in the GCR flux and spectral index corresponding to solar rotation, especially near sunspot minimum, while 
\add{the} flux
occasionally exhibited a strong harmonic at 13.5 days, and that the magnetic field component along a nominal Parker spiral (i.e., the magnetic sector structure) is a strong determinant of such spectral and flux variations, with the solar wind speed exerting an additional, nearly rigidity-independent influence on flux variations.
Our investigation affirms the capability of ground-based NM stations to accurately and continuously monitor cosmic ray spectral variations in the long-term future.
 \end{abstract}


\section{Introduction} \label{sec:intro}

When cosmic ray ions 
with energy $\gtrsim1$ GeV interact in the Earth's atmosphere, they create showers containing various sub-atomic particles. 
The flux of these secondary particles, including neutrons, has been continuously monitored by ground-based detectors including neutron monitors (NMs) \citep{Simpson48,Hatton64} across 
\add{time} scales
\add{up} to decades \citep{Usoskin05}. 
The primary cosmic ray ions are mostly Galactic cosmic rays (GCRs), which traverse extensive regions of the heliosphere before reaching Earth.
Accordingly, measurements of the temporal variation of the GCR spectrum (i.e., the flux as a function of particle energy or rigidity $R=pc/q$, related to the particle's momentum per unit charge) can provide distinctive insights into particle transport conditions in the heliosphere as influenced by the solar wind speed and magnetic field and their fluctuations, which in turn are influenced by the solar activity cycle, solar storms, shocks, rotating solar wind streams, and their interactions. 
The flux variations over short time scales can be of the order of 1\%, so large detectors are required to provide sufficient statistical precision.

Ground-based cosmic ray detectors 
monitor variations in the cosmic ray flux above a certain threshold \add{or} ``cutoff'' with a detector-specific yield function \citep{Clem00}.
For NMs, which are mostly sensitive to atmospheric secondary neutrons from cosmic ray showers, there are two types of cutoffs.  
The geomagnetic cutoff relates to a charged particle's rigidity, which determines its trajectory in Earth's magnetic field \add{regardless} of the particle species.  
Locations near Earth's geomagnetic equator, and especially Southeast Asia, have the highest cutoff rigidity, up to 17 GV \citep{Smart09,Ruffolo20}, while locations near Earth's poles have a geomagnetic cutoff below 1 GV.
Polar NMs are affected by the other type of cutoff, an  atmospheric cutoff at $\sim$1 GV, related to the kinetic energy per nucleon needed to generate a detectable flux of secondary particles at ground level \citep[e.g.,][]{Mangeard16-latsur}.
Historically, GeV-range spectral variations were tracked over monthly or yearly time scales by comparing the fluxes from NMs at different cutoff rigidities \citep{Usoskin05} or based on repeated balloon flights \citep{Abe16}.
Simple estimates of spectral variation based on the count rate ratio between two NMs at different cutoffs were found to be unreliable over shorter time scales, presumably due to different systematic effects \citep{Ruffolo16}. 

New techniques have recently been developed for tracking variations of the GeV-range GCR spectrum.
Special electronics can record the distribution of the time delay between successive neutron counts in a NM, in effect measuring the neutron multiplicity \citep{BieberEA04}, which relates to the \add{energy of secondaries from cosmic-ray showers} and in turn the primary cosmic ray energy.
From such distributions, the effect of chance coincidences \add{between neutron counts associated with different cosmic ray showers} can be removed to extract the leader fraction, $L$, i.e., the fraction of \add{neutron counts that did not follow a previous neutron count} from the same cosmic ray shower, which represents the inverse multiplicity \citep{Ruffolo16}.  
The leader fraction can serve as a proxy of the spectral index, as demonstrated by an analysis of data from a ship-borne NM for latitude surveys during 2000-2007 \citep{Mangeard16-latsur} and allows a measurement of spectral variations over time scales as short as days \citep{Ruffolo16} or as long as a decade \citep{Banglieng20}.
In principle these data can be made available in near-real time.

Even more recently, daily direct measurements from the AMS-02 detector on the {\it International Space Station} (ISS) have been published, including spectra of protons \citep{Aguilar21} and helium nuclei \citep{Aguilar22} during 2011-2019.  
These direct measurements have the advantages of identification of the primary cosmic ray species and accurate measurement of the energy.
AMS-02 can be expected to continue operation along with the ISS at least through the year 2030.
Another recent innovation is the refinement of the comparison of historical NM count rates from different cutoff rigidities, using several stations and paying careful attention to systematic effects on individual NM rates.  
With assumptions of a specific interstellar cosmic ray spectrum and the force-field model of solar modulation (i.e., of solar effects on GCRs), \citet{Vaisanen23} provide a daily tabulation of the modulation potential $\phi$ during 1964-2021 that expresses the long-term and short-term spectral variation of GCRs.  
However, it should be noted that a solar magnetic polarity reversal occurs near each solar activity maximum, which causes charge-sign-dependent solar modulation \citep{GarciaMunoz86,Adriani16,Aguilar23} and the ``crossover'' effect when comparing spectra for the same particle population from times of opposite polarity \citep{Moraal89,Nuntiyakul14,Poopakun23}, while the force-field model assumes a standard functional form and precludes such effects.

The interstellar spectrum is believed to be close to a power law over about 10-1000 GV, and as the rigidity decreases below $\sim$10 GV, the spectrum rolls downward \citep{Ghelfi16}.
Due to inhomogeneity and structures in the interplanetary magnetic field, as GCRs enter the heliosphere they experience scattering and mirroring in their motion along the large-scale magnetic field and also undergo perpendicular diffusion and drift motions \citep{Engelbrecht22}, which can collectively be called solar modulation \add{\citep{Potgieter13}}.
This causes the GCR spectrum to roll down further at lower rigidity, with quasi-periodic variations across a range of timescales. 
When the sunspot number is higher, on average there are more frequent solar storms and the solar wind has a higher speed and a stronger mean magnetic field strength and magnetic fluctuations.  
All of these features serve to further inhibit the access of GCR to the inner heliosphere, especially at lower rigidity, resulting in an anti-correlation with the sunspot number according to the 11-year solar activity cycle \citep{Forbush54}.  
There is also an effect of solar magnetic polarity, e.g., on large-scale drift motions in the heliosphere, resulting in spectral variation with the roughly 22-year solar magnetic cycle \citep{Jokipii81,Moraal89,Nuntiyakul14}.
Variations in the GCR spectrum also include the 27-day variation \citep{Fonger53} and its harmonics, Forbush decreases due to solar storms \citep{Forbush37}, and diurnal variation \citep{Hess36} and its harmonics due to Earth's rotation in the presence of cosmic ray anisotropy. 

The variation with a period of roughly 27 days, i.e., the synodic solar rotation period \add{relevant to the recurrence of solar wind} structures, is stronger near sunspot minimum and is associated with heliolongitudinal asymmetry in the solar wind. 
For example, at some times there are one or more coronal holes that generate equatorial high speed streams in the solar wind along certain heliographic longitudes.  
(Note that Earth's orbital plane is tilted only 7$^\circ$ from the Sun's equatorial plane.)
As scattering centers \add{(magnetic fluctuations)} convect with the solar wind, the solar wind speed near Earth may directly affect the GCR flux, and an antiphase correlation \add{between these} has been reported \citep{Modzelewska19}.
Furthermore, as the Sun rotates a high speed stream may rotate to a longitude previously occupied by a slower stream, leading to an interaction region that can generate shocks; such corotating interactions (CIRs) may also explain a periodic modulation of the cosmic ray spectrum  \citep{Moses87,Ghanbari19}.
Furthermore, the solar wind has a magnetic sector structure, with a magnetic field predominantly toward or away from the Sun according to the polarity of the solar source region,
a pattern that tends to repeat with the 27-day rotation period.
The heliospheric current sheet (HCS) is found at the boundary between magnetic sectors. \add{  The HCS has been observed to act as a barrier to low-energy energetic particle transport \citep{Droege90},
while in the GeV-range there is a significant drift along the HCS, the direction of which depends on the polarity of the preceding and following magnetic sectors; this drift process is believed to play a major role in solar modulation as a whole \citep{Jokipii81}.}
When examining the amplitudes of 27-day variations during sunspot minimum periods of different magnetic polarity, \citet{Modzelewska21a} found similar amplitudes in spacecraft observations of cosmic rays of $<1$ GeV, which they attributed to a dominant effect of turbulent diffusion, but different amplitudes for more energetic cosmic rays observed by NMs, which was attributed to the large-scale drift effect.


\begin{figure}[t]
	\centering
	\includegraphics[scale = 0.5]{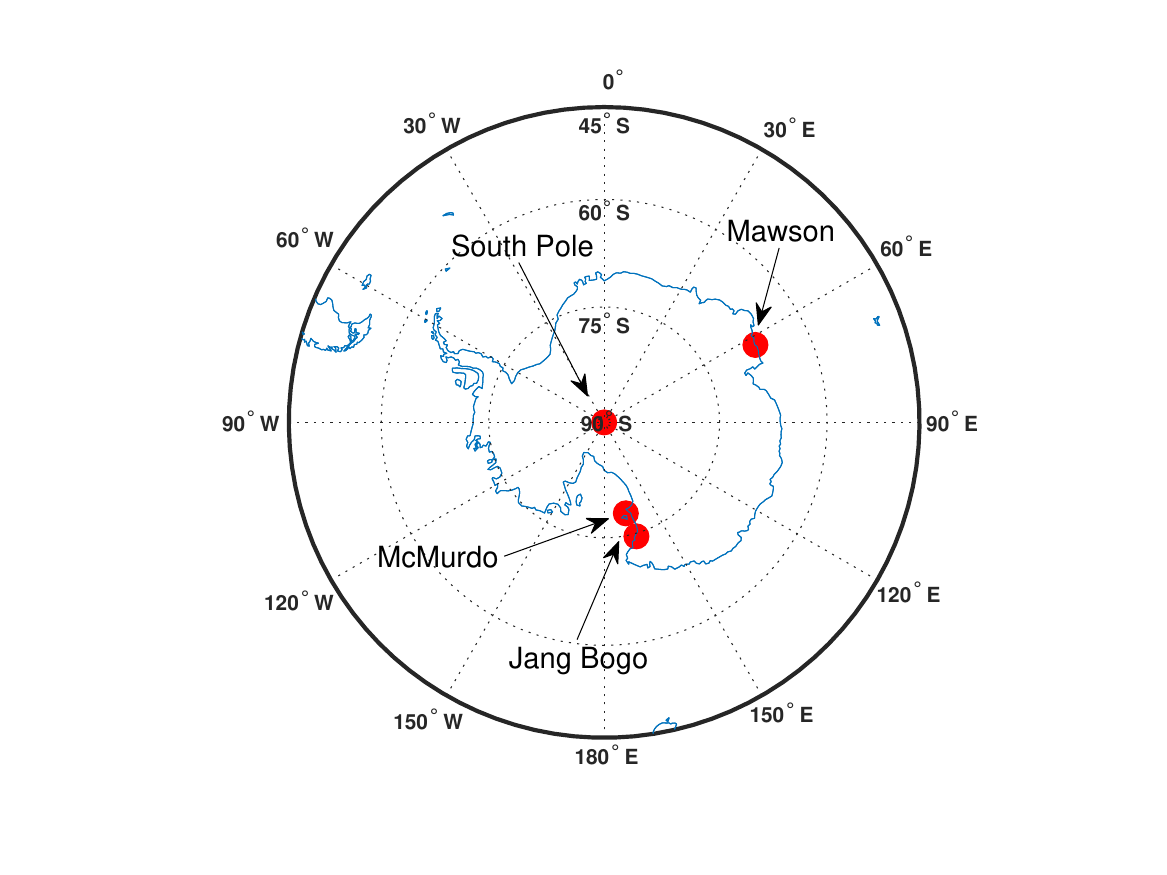}
	\caption{Map of Antarctica indicating the locations of NMs used in this work (red dots): the South Pole station at an altitude of 2820 meters; and McMurdo, Jang Bogo, and Mawson stations near sea level.  
 In each case, the rigidity-dependent response to primary cosmic rays has an atmospheric cutoff (i.e., threshold) at $\sim$1 GV.}
	\label{AntMap}
\end{figure}

The longest record of leader fraction data (2007-present) is from the Princess Sirindhorn Neutron Monitor at 2560 m altitude at Doi Inthanon, Thailand \citep{Ruffolo16,Banglieng20}, the fixed station with the world's highest vertical cutoff rigidity, $\sim$17 GV.  
In the present study, we provide complementary leader fraction information since 2015 from four Antarctic NMs, with several qualitatively different features\add{;} preliminary reports were presented by \citet{Muangha21,muangha23}.
One is that these polar NMs near sea level have atmosphere-limited cutoffs at $\sim$1 GV 
and therefore have similar response functions, so their variations in $L$ should be directly comparable, \add{although} the absolute value of $L$ may depend on the electronic dead time for time delay data, the software and firmware versions used for data acquisition, etc. 
Another feature is that the Antarctic stations are quite dry so there is no need to correct $L$ for atmospheric water vapor; at Doi Inthanon, the need for such a correction dominates the uncertainty in $L$.
Furthermore, GCR spectral variations due to solar modulation are much stronger at lower rigidity, making them easier to measure accurately for both NMs and AMS-02.

Here we have compared $L$ from the South Pole (SP) NM with the available AMS-02 data during their period of overlap, 2015-2019, confirming that it provides a reliable and accurate measure of the cosmic ray spectral index.  
The specialized electronics for accumulating neutron time delay distributions have also been deployed to three other NMs in Antarctica \add{at} McMurdo, Jang Bogo, and Mawson stations; see Figure \ref{AntMap}. 
We expect them to exhibit a similar energy-dependent response to primary cosmic rays, and checked for consistent variations in $L$ among these three monitors\add{, noting} that the response of the SP NM is somewhat different because of the higher altitude and lower atmospheric depth at the South Pole.
Then a wavelet analysis was used to describe periodic variations during 2015-2023 in the SP NM count rate and leader fraction, and also in relevant properties of the solar wind: its speed, magnetic field magnitude, and magnetic field component along the Parker spiral direction as a measure of the sector structure.
This allows us to draw conclusions as to which of the physical processes described above can explain the observed periodic variations in the cosmic ray flux and spectral index.

\section{Neutron Time-Delay Measurements} \label{sec:Data}

\subsection{Neutron Monitor Data}

Neutron monitors (NMs) are ground-based detectors of secondary particles produced in atmospheric cascades from primary cosmic ray ions. 
They consist of several proportional counters filled with $^{10}$BF$_3$ or $^3$He gas, which are sensitive to the neutrons produced by the interaction of the secondary particles with a dense lead producer. 
About 80\% of the count rate is due to atmospheric secondary neutrons of energy $>10$ MeV \citep{Mangeard16-PSNM}.
Each interaction typically produces $\sim$6 neutrons of MeV-range energy, which are further moderated in the polyethylene moderator immediately surrounding the proportional counters and/or the polyethylene reflector that surrounds the monitor.
The key function of an NM is to register the count rate of neutron capture events in the proportional counters.

The NM at the US Amundsen-Scott South Pole station, at an altitude of 2828 m, has three 1NM64 detectors (i.e., each is a detector of NM64 design with one proportional counter) on an outdoor platform.  
Each proportional counter and its surrounding moderator, lead, and reflector are in a separate insulated, heated box with a regulated temperature. The proportional counter tubes are filled  with $^3\text{He}$ gas to detect neutrons via the induced fission reaction $n + ^3\!\mathrm{He} \rightarrow p + ^3\!\mathrm{H}$, which produces kinetic energy $Q=764$ keV. The SP NM started using electronics to record the time-delay histograms on one NM tube in 2013 December, and useful data were recorded until 2014 November. 
Then electronics with newer firmware were installed on the complete set of 3 counter tubes in 2015 March. 
Because of an uncertain relative calibration between $L$ from the two sets of electronics \citep{Banglieng20}, here we only analyze data starting in 2015.

The South Korean Jang Bogo (JB) Station is located at latitude 74.6\textdegree{}S, longitude 164.2\textdegree{}E, and altitude 29 m. The JB NM is effectively a 6$\times$3NM64 neutron monitor, with a reflector almost completely surrounding each of six sets of three $^{10}\mathrm{BF}_3$ gas counters, which detect neutrons via the induced fission reaction $n + ^{10}\!\mathrm{B} \rightarrow ^4\!\mathrm{He} + ^7\!\mathrm{Li}$, which produces kinetic energy $Q=2.31$ or 2.79 MeV. 
The neutron monitor that is now at JB station was moved there in stages from the US McMurdo (MC) Station. 
The JB station and the MC station are quite close to each other geographically, at a distance of about 360 km. 
Specialized electronics that record time delay distributions were installed after the first stage of the move, providing time delay histograms from 6 of the 11 remaining counters in the MC NM and all 5 functioning counters at JB \citep{Jung16}.
The remaining 12 counters (11 NM counters and 1 bare counter without surrounding lead or polyethylene) at MC were removed in 2017 January and installed at JB in 2019 January.
Since then, JB time delay histogram data have been recorded from 11 counter tubes, but such data from one of the tubes have been unreliable, so we used $L$ from 10 tubes in this analysis. 

The NM at the Australian Mawson (MA) Station, at latitude $67.6$\textdegree{}S, longitude $62.8$\textdegree{}E, and altitude 30 m, has a 3$\times$6NM64 configuration with $^{10}\mathrm{BF}_3$ gas counters.  The MA NM was upgraded in 2020 February to record time-delay histograms from all 18 counter tubes, though one of the counters stopped working in 2020 September.


\subsection{Extracting the Leader Fraction and Correction for Atmospheric Pressure}
\label{sec:Extract_L}
A specialized electronic data acquisition system has been developed to record hourly distributions of time interval between consecutive neutron captures in the same counter (as used in this work), hourly distributions of the interval between an event on one counter and the next event on a second counter \citep{Saiz17,Saiz19}, and the absolute timing of each neutron detection on every counter for selected events \citep{Evenson22}. 

The leader fraction, $L$, is a statistical measure representing the proportion of neutron counts that do not follow another count in the same tube, stemming from the same primary cosmic ray. 
More energetic primary cosmic rays give rise to more energetic secondary particles, leading to the production of more neutrons interactions within the neutron monitor (typically in the lead producer). Consequently, these neutrons manifest as sequences of multiple counts in the same tube with shorter time delays. 
Thus the cosmic ray spectrum's hardness, characterized by a lower spectral index, contributes to a low value of the leader fraction.
\add{In addition, $L$ can be identified as an inverse multiplicity, where the multiplicity is the number of neutron counts from one counter tube that were associated with the same cosmic ray shower.}


At long time delay $t$, a time-delay distribution exhibits an exponential tail \add{\citep[see Figure 3 of][]{Ruffolo16}}, indicative of the random time delay between 
neutron counts associated with different primary cosmic rays.  
The tail can be fit by $n(t) = Ae^{-\alpha t}$ over the range 5 ms $<t<t_\mathrm{max}=$ 142 ms, where $n(t)$ is the probability density (corrected for missing data at $t>t_\mathrm{max}$) and $A$ and $\alpha$ are fit parameters.
In contrast, at short time delays, the distribution has a strong excess over that exponential function. 
This non-exponential behavior is attributed to temporally associated neutron counts originating from the same primary cosmic ray events \citep{BieberEA04}. 
For each hourly histogram from each counter, we determined $L$ from fit parameters of the exponential tail following \citet{Banglieng20}: $L = A/(\alpha e^{\alpha t_d})$ where $t_d$ is the effective electronic dead time.
$L$ is also normalized to account for changes in the data acquisition system and physical configuration; see Appendix A for details.

NMs are very stable and are suitable for studies of long-term solar modulation of cosmic rays. However, both the NM count rate $C$ and leader fraction $L$ must be corrected for changes in atmospheric pressure, which is a proxy for air mass, due to absorption of secondary particles in the atmosphere. 
While $C$ is anti-correlated with pressure, $L$ has a positive correlation \citep{Ruffolo16}. To account for atmospheric pressure variations, we employ a standard barometric correction method: 
\begin{equation}
C_\textrm{cor} = C_\textrm{uncor}\exp[\beta(P - P_0)], 
\end{equation}
where $C_\textrm{cor}$ is the pressure corrected count rate, $C_\textrm{uncor}$ is uncorrected $C$, $\beta$ is a pressure coefficient, $P$ is pressure, and $P_0$ is the reference pressure that has historically been used for each station.
Values of $(\beta, P_0)$ in units of \% hPa$^{-1}$ and hPa, respectively, are (0.74\%, 680) for SP, (0.74\%, 973.25) for MC, (0.74\%, 978) for JB, and (0.708\%, 990) for MA.

We use a similar formula for pressure correction of the leader fraction: 
\begin{equation}
L_\textrm{cor} = L_\textrm{uncor}\exp[-b(P - P_0)],
\end{equation}
where the pressure coefficient for $L$ is expressed as $b = -\beta$ to stress the positive correlation of the measured (uncorrected) $L$ with atmospheric pressure.
The determination of the pressure coefficient
applies a linear fit to $\Delta \ln L$ versus $\Delta P$, where ``$\Delta$'' indicates the deviation from a 13-day running average. 
It is noteworthy that different $b$ values are obtained for different data acquisition software versions and for different counter tubes,  depending on a tube's position and electronic dead time.  Mean values of $b$ are $0.0187\,\%\, \text{hPa}^{-1}$ for SP, $0.00934\,\%\, \text{hPa}^{-1}$ for MC, $0.00865\,\%\, \text{hPa}^{-1}$ for JB, and $0.00821\,\%\, \text{hPa}^{-1}$ for MA.
Because atmospheric pressure can vary significantly from one hour to the next, we applied the pressure correction to $L$ for each hour and each counter; then $L$ was averaged over all counters in the NM to obtain an hourly $L$ value.
For improved statistics, these hourly values are usually combined into daily averages.

\section{South Pole Leader Fraction Compared with AMS-02 Spectral Index} \label{sec:Spect}
To establish a meaningful comparison between the SP $L$ and the cosmic ray spectral index determined by direct space-based measurements, we calculated the spectral index $\gamma$ through a power-law fit, utilizing the daily proton flux as a function of rigidity, $j(R)$, within a specified rigidity range.  
In other words, $j(R)$ is fit to $j_0R^{-\gamma}$ with $j_0$ and $\gamma$ determined as fit parameters. 
Useful data for this analysis are available from the AMS-02 detector, which provides measurements of daily cosmic ray proton fluxes for rigidity intervals ranging from 1 to 100 GV, from the start of our SP $L$ data set, 2015 March 3, up until October 29, 2019 \citep{Aguilar21}. 
We found a particularly strong correlation between SP $L$ and the AMS-02 spectral index when the latter was determined by fitting the measured spectrum over the rigidity range of 2.97 to 16.6 GV.

\begin{figure}[t]
	\centering
	\includegraphics[scale = 0.4]{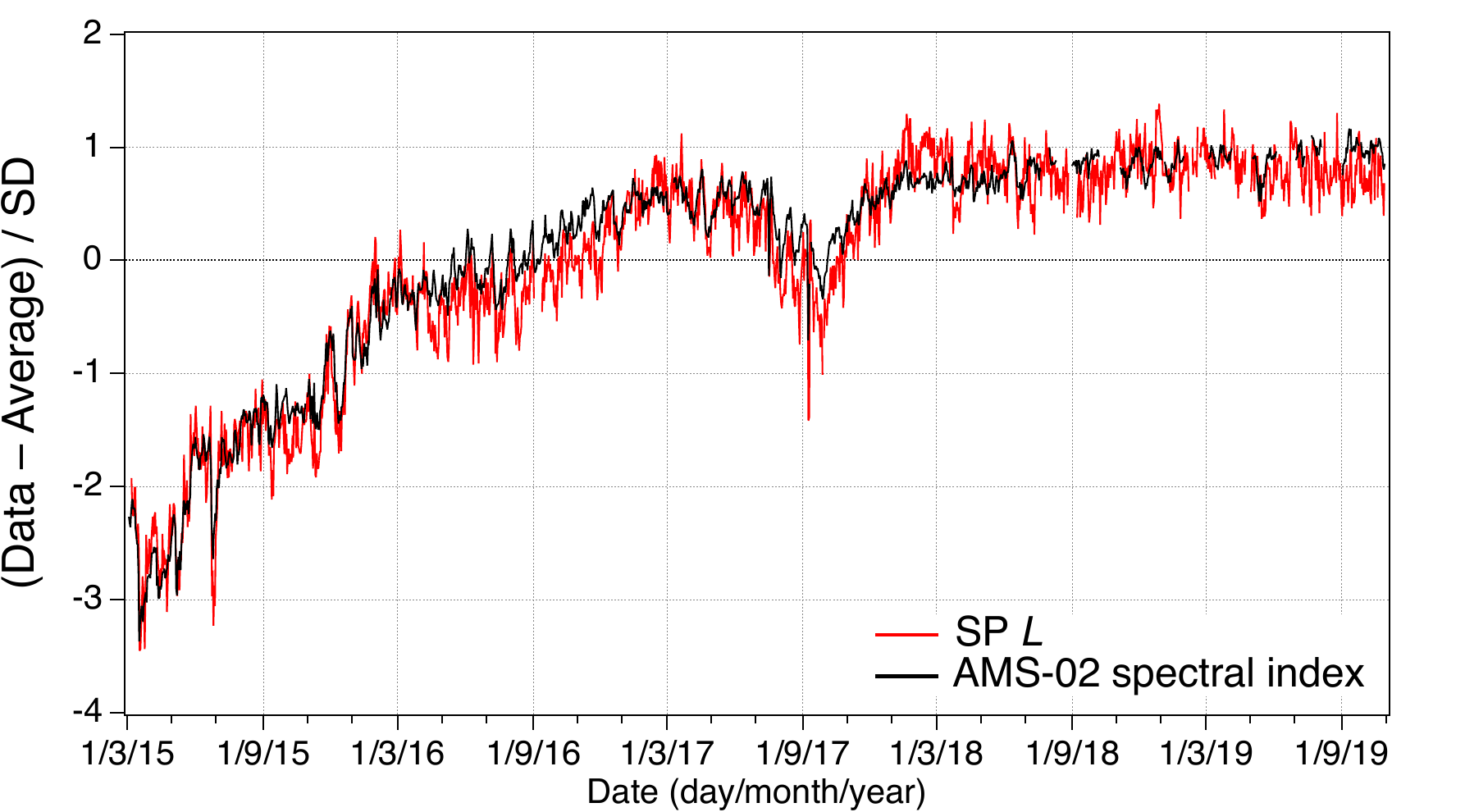}
	\caption{Normalized daily values of the South Pole leader fraction
 (red) and AMS-02 proton spectral index (2.97--16.6 GV, black) from 2015 March 3 to 2019 October 29 when both SP $L$ and AMS-02 proton data were available, confirming that $L$ serves as a proxy of the cosmic ray proton spectral index.}
	\label{fig.SPL_ID_d}
\end{figure}

For a more effective comparison of variations, both SP $L$ and the AMS-02 proton spectral index were \add{standardized} using the formula (daily data - average)/(standard deviation). The resulting time series, plotted in Figure \ref{fig.SPL_ID_d}, confirm a close relationship between variations in the two parameters, with some minor differences. 
In general, SP $L$ exhibits greater 
\add{short-term} fluctuation compared to the AMS-02 proton spectral index.
Furthermore, SP $L$ occasionally exhibits a minor systematic deviation from the AMS-02 proton spectral index, exhibiting a lower value during the austral summer during some but not all years. 
These fluctuations may represent a seasonal atmospheric effect near the South Pole; indeed, the polar atmosphere is systematically different during the winter (without sunlight) and summer (with sunlight).
Nevertheless, these seasonal variations are much weaker than the effect of solar modulation and no correction for seasonal atmospheric effects (other than atmospheric pressure) are applied in this analysis.


\begin{figure}[t]
	\centering
	\includegraphics[scale = 0.3]{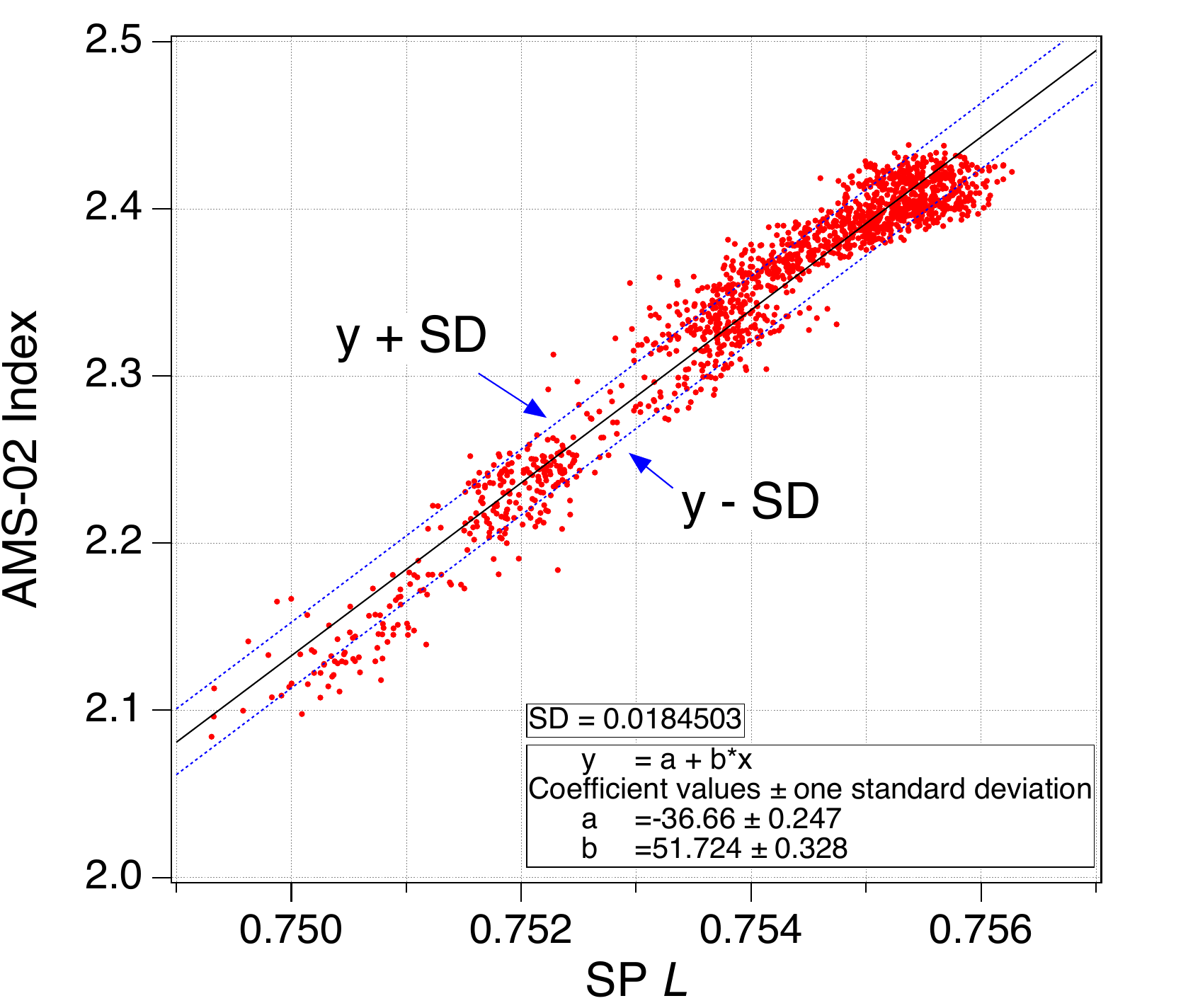}
	\caption{Daily AMS-02 proton spectral index over the rigidity range 2.97--16.6 GV versus the South Pole (SP) leader fraction $L$. The black line represents the linear fit to the data.
 The blue dashed lines, displaced from the black line by one standard deviation,  illustrate the associated uncertainty when employing $L$ as an estimator of the spectral index. 
 }
	\label{fig.AMS02vsL}
\end{figure}

The relationship between the AMS-02 proton spectral index and SP $L$ (Figure \ref{fig.AMS02vsL}) over the entire data set from 2015 March to 2019 October, ranging from near sunspot maximum to near sunspot minimum, can be approximated as a linear relationship. 
We can convert SP $L$ into an estimate of the proton spectral index $\gamma$ over 2.97--16.6 GV  using that relationship, 
\begin{equation}
\gamma = -36.66 + 51.72 L.
\label{eq:linear}
\end{equation}
To evaluate the accuracy of the leader fraction's prediction of the spectral index, we computed the difference between the predicted and measured $\gamma$.
This difference was utilized to calculate the residual standard deviation of 0.018, which can be considered as the uncertainty in estimating the spectral index using SP $L$. 
This includes both statistical uncertainty and systematic deviations from a linear relationship.
This low overall uncertainty implies that the leader fraction provides an accurate estimate of the cosmic ray spectral index.

\begin{figure}[t]
	\centering
	\includegraphics[scale = 0.4]{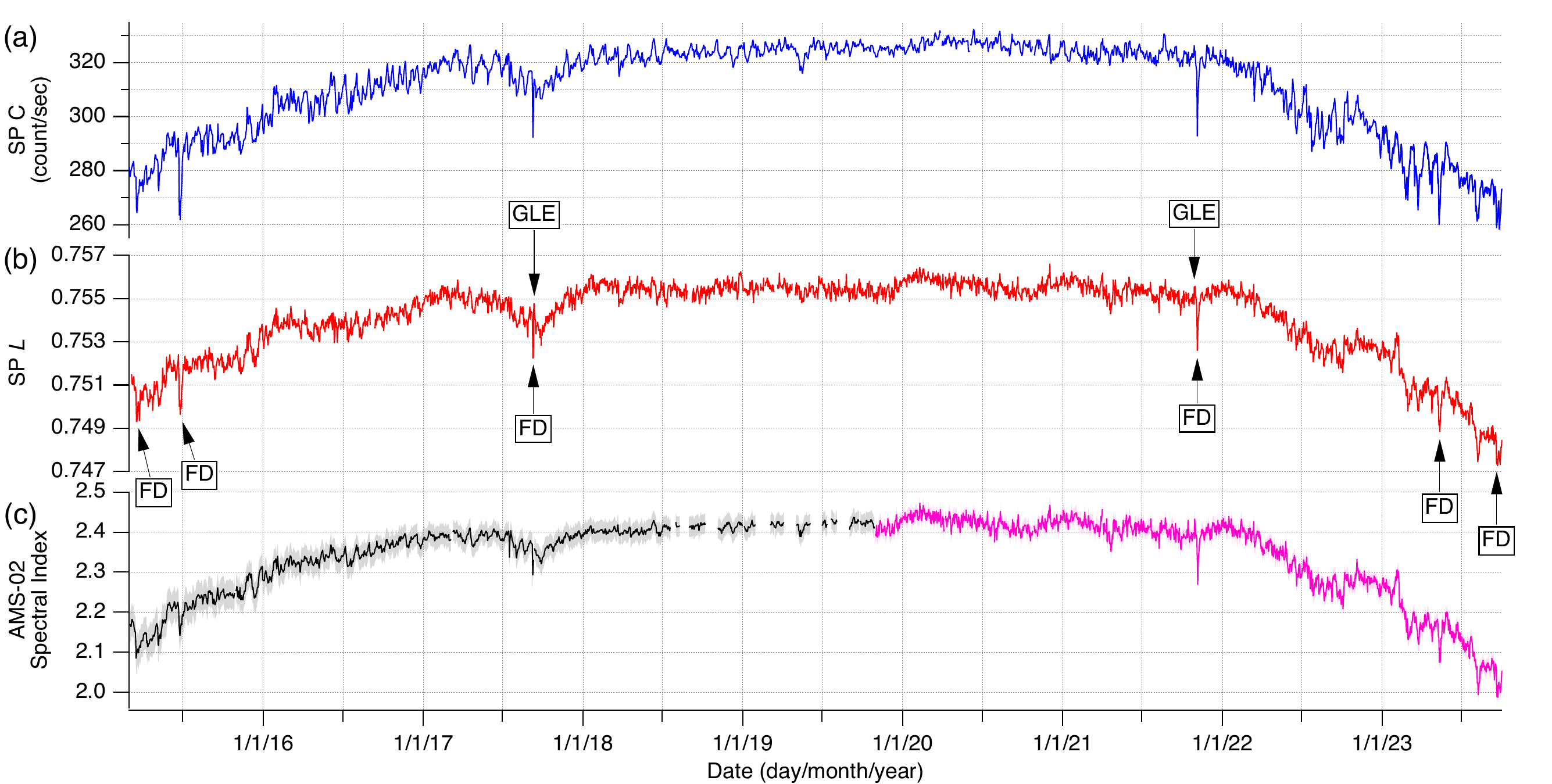}
	\caption{Time series of daily data from 2015 March to 2023 September: (a) South Pole (SP) neutron monitor count rate ($C$), reflecting changes in the cosmic ray flux at rigidities $\gtrsim$ 1 GV.  (b) SP  leader fraction $L$, a measure of the cosmic ray spectral index $\gamma$.  Labels ``FD'' and ``GLE'' indicate Forbush decrease and ground level enhancement events, respectively, in association with solar storms. (c) AMS-02 proton spectral index over the rigidity range 2.97--16.6 GV\@. The black line is based on measurements from 2015 March 3 to 2019 October 29. Grey shading around the line indicates the uncertainty associated with the fitting process. The magenta line represents the prediction of the spectral index beyond 2019 October 29, based on SP $L$ using Equation (\ref{eq:linear}), i.e., our prediction for what will be found from AMS-02 data when they are made public.  
 Shading around this line represents the corresponding uncertainty. 
}
	\label{fig.SPC_SPL_AMS02}
\end{figure}

Figure \ref{fig.SPC_SPL_AMS02} presents time series of daily values of SP $C$ and $L$ from 2015 March to 2023 September.
Overall, the GCR count rate and spectral index display similar trends as enhanced solar modulation causes the spectrum to roll down more strongly at lower rigidity;
thus a lower flux is generally associated with a flatter spectrum and lower spectral index \citep{Banglieng20}. 
The label ``FD'' refers to a \add{transient} Forbush decrease in the GCR spectrum due to passage of an interplanetary shock and/or coronal mass ejection from a solar storm.
Note that different FD events have different levels of spectral change relative to the flux change \citep{Ruffolo16}.  
The label ``GLE'' refers to a ground level enhancement, a solar radiation storm in which the flux of \add{energetic} solar particles is so strong that it can be observed as an enhancement over the GCR flux in ground-based NM count rates \add{\citep{Poluianov17}}.
The two GLE events during this time period produced flux increases that were so small that they are not visually evident in the long-term time series of $C$, yet they are visible in the time series of $L$ because the spectrum of solar particles is much steeper than that of GCRs \citep[see also][]{Banglieng20}.  
Figure \ref{fig.SPC_SPL_AMS02}(c) shows the AMS-02 proton spectral index $\gamma$ over the rigidity range 2.97--16.6 GV, which closely tracks the time series of SP $L$. 
[Note that data with relativistic solar particles from the GLE of 2017 Sep 10 were not included in the published AMS-02 data.]
The magenta line represents our prediction, based on SP $L$ using Equation (\ref{eq:linear}), for the spectral index $\gamma$ that \add{can be expected} from AMS-02 data beyond 2019 October 29 when those data are made public.
From panel (c) we see that $\gamma$
had an overall increase from  $\approx$2.1 in 2015, near sunspot maximum, to $\approx$2.4 in 2019, near sunspot minimum. 
Similarly, SP $L$ mostly increased from 2015 to 2019 and then decreased again starting in 2022 as the sunspot cycle heads toward another maximum.

\begin{figure}[!ht]
	\centering
	\includegraphics[scale = 0.41]{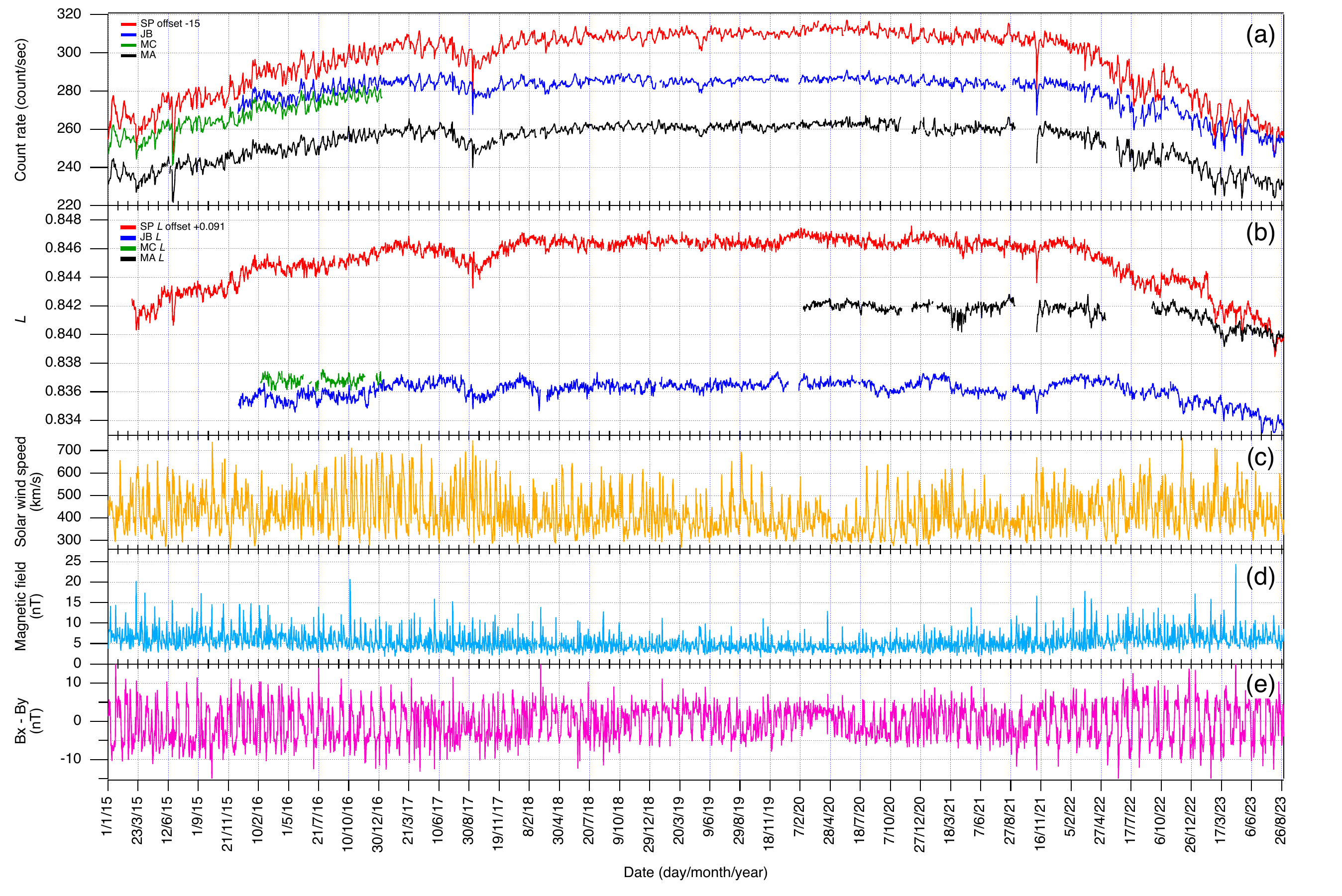}
\caption{Time series of daily data from 2015 January to 2023 September: (a) Count rate $C$ and (b) leader fraction $L$ from four Antarctic NMs, at SP (red), MC (green), JB (blue), and MA (black). (c) Solar wind speed (yellow), (d) $B$ magnitude (blue) and (e) $B_x - B_y$. 
There are vertical grid lines every 81 days and minor ticks along the axis every 27 days. \add{Note that for ease of comparison, SP $C$ data have been shifted downward by 15 s$^{-1}$ and SP $L$ data have been shifted upward by 0.091.}}
	\label{fig.L_C_solarp}
\end{figure}

\section{Comparison among Leader Fractions from Four Antarctic Neutron Monitors} \label{sec:Compare_Ant_L}

The four Antarctic NM stations considered here all have geomagnetic cutoffs so low that their count rate responses as a function of primary cosmic ray rigidity are limited by the atmospheric cutoff rigidity.
A key difference is that SP is at 2820 m altitude while MC, JB, and MA are near sea level.  
Because of the altitude difference, SP NM observes a much higher count rate per counter of secondary particles from cosmic ray showers as well as a higher multiplicity and lower leader fraction, which is consistent with the positive pressure coefficient noted in Section \ref{sec:Extract_L}. 
Despite having only three neutron counters, the South Pole NM recorded the highest daily count rate $C$, ranging from 258 to 332 s$^{-1}$, while concurrently displaying the lowest 
daily leader fraction, $L$, ranging from 0.747 to 0.756 (see Figure \ref{fig.L_C_solarp}).

We observe a strong correlation between the count rates at JB and MC NM stations during the period of simultaneous operation from 2015 December to 2017 January \citep[see also][]{Kittiy23} because these two stations are close to each other, and indeed their time variations are quite similar to those at MA (which is also near sea level).
The different levels of $C$ and $L$ from MA relative to JB can mostly be attributed to its higher reference pressure $P_0$ (see Section \ref{sec:Extract_L}). 
When we correct the data to the same $P_0$, the $C$ and $L$ values from JB and MC are much closer.  
In any case, the absolute $C$ and $L$ values from an NM depend on the number of operating counters and their configuration, electronic dead times, computer and software version (for $L$), and surrounding structure and environment, so NM data are generally interpreted in terms of variations rather than absolute values. 
Figure \ref{fig.L_C_solarp} shows a reasonable consistency in the variations in $L$ and especially in $C$ among the three sea-level stations, while SP $C$ and $L$ exhibit stronger variation over the sunspot cycle as well as stronger shorter-term variations because the high-altitude SP NM as a lower atmospheric cutoff and is relatively more sensitive to primary cosmic rays at lower energy.


\section{Periodic Variations in the Cosmic Ray Flux and Spectral Index}\label{sec:Periodic_L}
In this section, we consider periodic variations in the cosmic flux, as indicated by $C$, and spectral index, as indicated by $L$, from these four Antarctic NMs.  
Because our data set spans the years 2015 to 2023, which is shorter than an 11-year sunspot cycle, we focus on their shorter-term variations. 
The 27-day variation is particularly prominent in the time series of $C$ and $L$ as shown in Figure \ref{fig.L_C_solarp}.  
In addition, this figure shows time series of some relevant 
solar wind parameters: the daily solar wind velocity, magnetic field magnitude ($B$), and the difference in magnetic field components along the GSE $x$- and $y$-directions, $B_x - B_y$.
The combination $B_x - B_y$ indicates a magnetic field component along a nominal Parker spiral field direction \citep[][]{Parker58} and therefore serves to indicate the magnetic sector structure of magnetic fields directed toward the Sun (for $B_x-B_y>0$) or away from the Sun \citep[for $B_x-B_y<0$][]{Buatthaisong22}.
These daily values of solar wind parameters were obtained from the {\it Advanced Composition Explorer} Real-Time Solar Wind
database: (\url{https://sohoftp.nascom.nasa.gov/sdb/goes/ace/}).
By examining these possible solar wind determinants, we aim to understand the key physical properties that underlie GCR variability, particularly for the $\approx$27-day period and its harmonics.

\begin{figure}
	\centering
	\includegraphics[scale = 0.43]{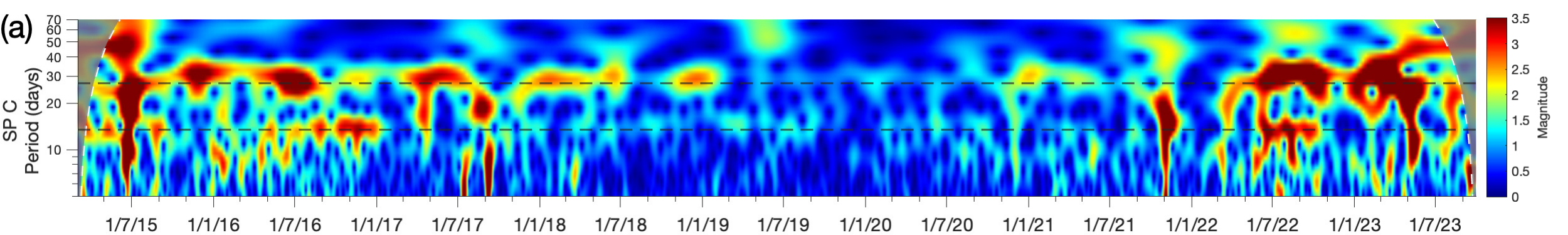}
	\includegraphics[scale = 0.43]{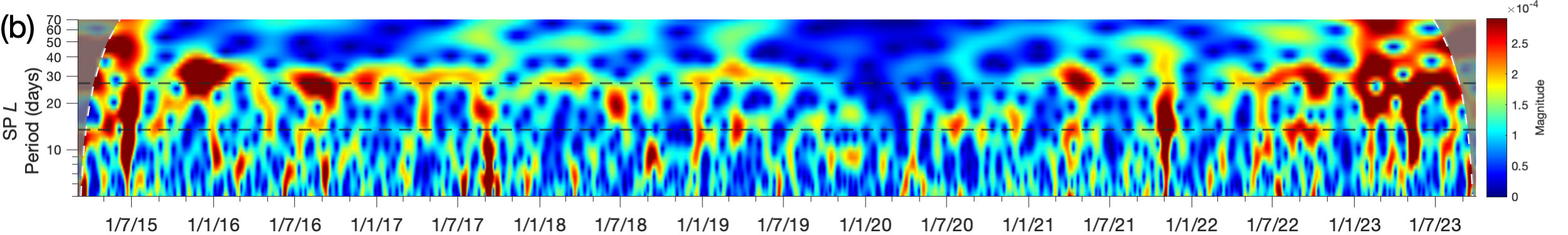}
        \includegraphics[scale = 0.43]{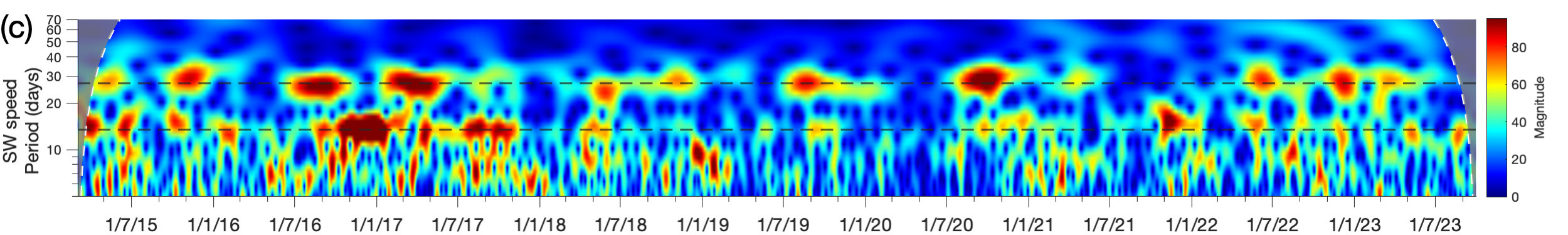}
	\includegraphics[scale = 0.43]{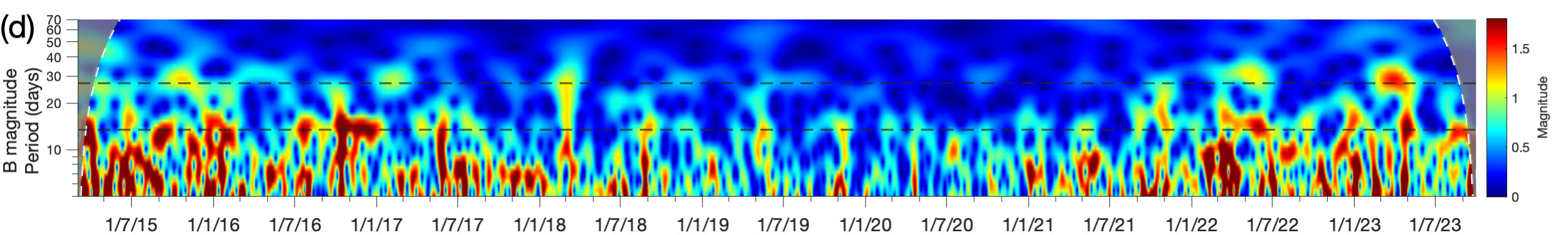}
	\includegraphics[scale = 0.43]{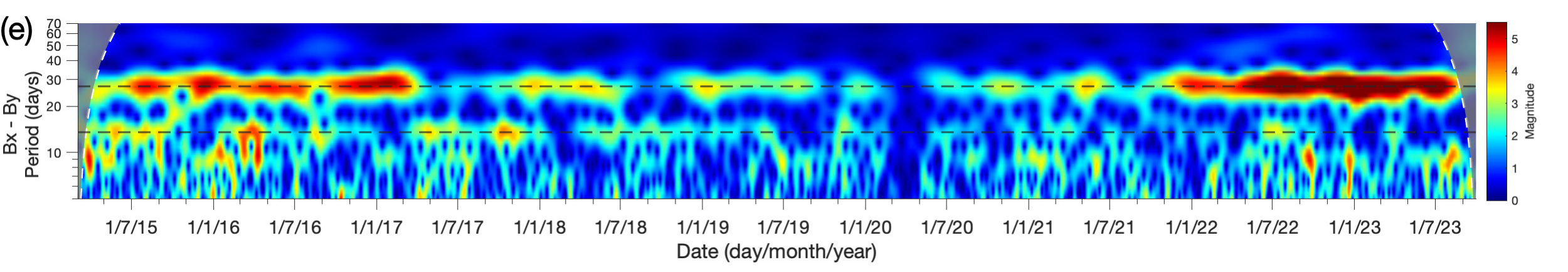}
\caption{Wavelet power as a function of period and time from 2015 March to 2023 September for daily values of (a) SP $C$ as a measure of cosmic ray flux at $\sim$10 GV, (b) SP $L$ as a measure of the cosmic ray spectral index at $\sim$10 GV, and solar wind parameters: (c) speed, (d) magnetic field magnitude $B$, and (e) $B_x - B_y$ as a measure of the magnetic sector structure. Horizontal black dashed lines in each panel represent the periods of 27 days and 13.5 days.
The white dashed line near the edge corresponds to the $95\%$ confidence level of the wavelet analysis. The color bar indicates the power of a period range from low to high power (blue to red).
$B_x - B_y$ is the solar wind parameter that best matches the SP $C$ and $L$ wavelet power, with dominant periodicity at the $\approx$27-day solar rotation period, only during times near solar maximum, and little power at the 13.5-day harmonic.  
We conclude that the magnetic sector structure is the key determinant of 27-day variation in the GCR spectrum at $\sim$10 GV and suggest drift along the HCS as the key physical process.
The solar wind speed also makes a nearly rigidity-independent contribution to the variation in $C$, or the cosmic ray flux, as evidenced by an occasional 13.5-day periodicity in both quantities.
}
	\label{fig.CWT}
\end{figure}

\begin{figure}
	\centering
	\includegraphics[scale = 0.25]{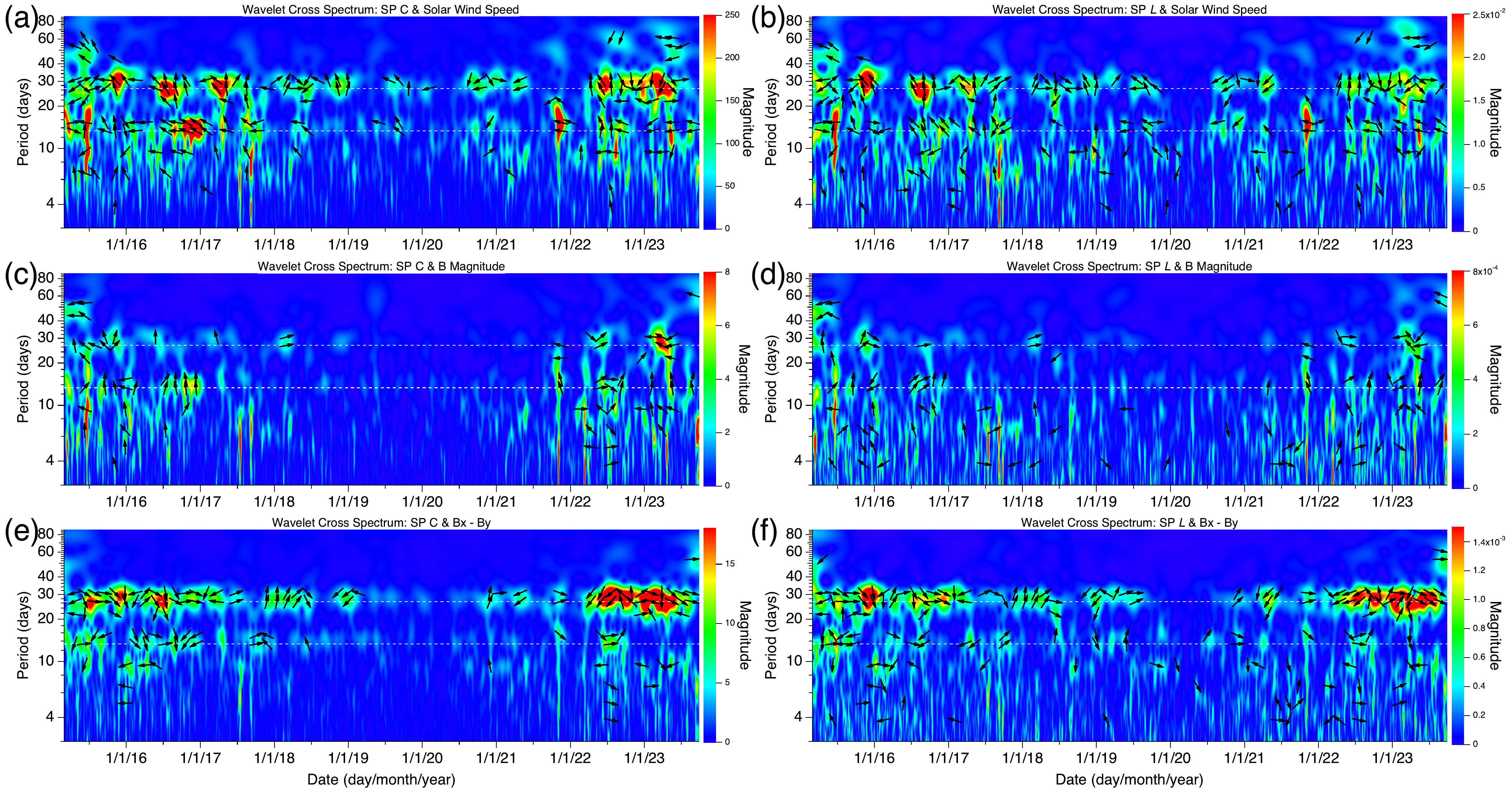}
\caption{\add{(Left) Cross wavelet spectrum between SP $C$ and solar wind parameters from March 2015 to September 2023: (a) solar wind speed, (c) $B$, (e) $B_x-B_y$. (Right) Cross wavelet spectrum between SP $L$ and solar wind parameters: (b) solar wind speed, (d) $B$, (f) $B_x-B_y$. Horizontal white dashed lines in each panel represent the periods of 27 days and 13.5 days. 
Where the magnitude exceeds \rev{20\% of the maximum magnitude}, the phase difference between the two quantities is indicated by the direction of an arrow, with zero phase difference indicated by an arrow to the right.}}
	\label{fig.CWS}
\end{figure}

To investigate the temporal changes in periodicity associated with the Sun's rotation, we compare the quasi-periodic variations of $C$, $L$, and heliospheric parameters using the Morlet-wavelet analysis (Figure \ref{fig.CWT}). We 
focus on results from the SP NM, which collected nearly continuous data since 2015 March. 
Because the median rigidity of the SP NM response to primary cosmic rays is at 11 GV \citep{Munakata22} and in Section \ref{sec:Compare_Ant_L} we showed that $L$ reflects the spectral index over 2.97--16.6 GV, roughly speaking one can say that SP $C$ and $L$ indicate the cosmic ray flux and spectral index, respectively, at $\sim$10 GV.

Both SP $C$ and $L$ exhibited much weaker periodic variations during 2018--2020, near sunspot minimum,  than during earlier or later years with a higher sunspot number and stronger solar activity.
This is also true for the magnetic field magnitude $B$ and for $B_x-B_y$, but not clearly the case for the solar wind speed, which sporadically indicates strong periodicity throughout the entire time period.  
This argues against the solar wind speed being the dominant determinant of variations associated with solar rotation in the GCR spectrum at $\sim$10 GV, which also applies to related phenomena such as CIRs, which occur when slower solar wind precedes faster solar wind.

The count rate $C$, representing the cosmic ray flux, exhibits strong periodicities at approximately 27 days and 13.5 days (i.e., the first harmonic of the solar rotation period).
\add{Note} that there are strong broadband signals in 2015 June, 2017 September, and 2021 October. These were times of strong solar storms and Forbush decreases.  These broadband signals reflect sudden changes that are physically distinct from the periodicity associated with solar rotation.
While the 27-day signal is frequently strong during times near solar maximum, the 13.5-day signal is strong in $C$ during only two time periods: in late 2016, when the 13.5-day signal temporarily replaced the 27-day signal, and in mid-2022, when both signals were particularly strong.  
Interestingly, the wavelet power for $L$ is strongly correlated in time with that for $C$ at the 27-day period, but almost never displays a strong signal at the first harmonic (at 13.5 days).
The same can be said of $B_x-B_y$.  
Indeed, among the solar wind parameters, $B_x-B_y$ provides the clearest match to the wavelet power map for SP $C$ and especially $L$.
Excluding the times with sharp Forbush decreases as noted above, the wavelet power patterns of $C$, $L$, and $B_x-B_y$ are all strong only during times near solar maximum (i.e., up to early 2017 and after late 2021) and are all dominated by the 27-day period, with a sporadic and much weaker signal at 13.5 days.
\add{The 27-day period of SP $L$ is most strongly associated with $B_x-B_y$, and the relationship with the magnitude $B$ is much weaker. 
Occasionally, the solar wind speed may exhibit some influence on $L$, although it is generally less significant.} 

\add{These trends are confirmed by the cross wavelet spectrum for each solar parameter, paired with $C$ and $L$, as depicted in Figure \ref{fig.CWS}.
The direction of the arrows indicates the phase relationship; arrows pointing to the right denote signals in phase or with zero delay.
Arrows are only shown where the magnitude of the cross wavelet spectrum (indicated by the color scale) exceeds \rev{20\% of the maximum magnitude}. In Figure \ref{fig.CWS}(a), it is evident that the 27-day variations in SP $C$ and solar wind speed mostly exhibited a phase difference between approximately $90^\circ$ to $260^\circ$ from March 2015 to September 2017 and from April 2022 to June 2023. Other time intervals show fluctuating phase differences, mostly in antiphase. Similar variations were observed for the 13.5-day period, but mostly in antiphase.}  
\add{During periods when SP $L$ and the solar wind speed exhibited high cross power, the observed phase difference fluctuated significantly. The temporal evolution of high power at the 27-day period in both SP $L$ and SP $C$ with $B_x-B_y$, as shown in Figure \ref{fig.CWS}(e) and (f), clearly exhibits a solar cycle dependence. Phase relationships sometimes occurred in antiphase and sometimes in phase, with $B$ exhibiting a weaker relationship with both SP $C$ and SP $L$. The phase between SP $L$ and solar wind speed fluctuates, indicating a variable phase relationship.}

Let us return to the two exceptional time periods when $C$ had a strong signal at the 13.5-day period.
During mid-2022, both $C$ and $L$ exhibited both 27-day and 13.5-day signals, though the 13.5-day signal is weaker.  
The origin of this dual-period pattern is unclear because none of the solar wind parameters exhibited a similar pattern.
Indeed this is the only time period when $L$ exhibited a clear enhancement in wavelet power at 13.5 days.
In contrast, during late 2016, while $C$ had a dominant 13.5-day signal, $L$ exhibited a 27-day period as usual.
Interestingly, at that time the solar wind speed and $B$ also exhibited a strong 13.5 day period.
Now $B$ cannot be said to closely represent the wavelet power for either $C$ or $L$, because it has almost no power at the 27-day period.
But the solar wind speed does exhibit a 27-day period most frequently, and during late 2016 it switched to 13.5 days just like $C$ did.  
Therefore it seems that the solar wind speed did play a role in determining variations in the cosmic ray flux during late 2016, even if $B_x-B_y$ seems to serve as a better determinant overall.
In contrast, the strong dominance of the 27-day period in $L$, specifically during times near solar maximum, corresponds closely to the pattern for $B_x-B_y$, which therefore can be considered as the best determinant of variation in the spectral index at $\sim$10 GV.

% ... 


\section{Discussion} 
In the present work we have further explored the use of a recently developed indicator of cosmic ray spectral variation, the NM leader fraction ($L$).  
Analyzing $L$ from four Antarctic NMs has allowed us to verify that $L$ exhibits similar variations for monitors with a similar response (especially for the sea level stations, MC, JB, and MA).  
The nearly continuous time series of $L$ from the South Pole NM since 2015 is shown to have a linear relationship with the spectral index inferred by fitting daily proton spectra over 2.97--16.6 GV from the space-based AMS-02 detector up to 2019 October, which were published recently.
This provides further confirmation \citep[see also][]{Mangeard16-latsur} that $L$ serves as a proxy for the spectral index.
We estimate an overall (statistical plus systematic) uncertainty of 0.018 in the spectral index as inferred from $L$. 


We have also performed a wavelet analysis to examine periodic variations in the South Pole NM's count rate $C$, which reflects GCR flux variations at $\sim$10 GV, and $L$, which reflects variations in the spectral index, during 2015 March to 2023 September, along with relevant solar wind parameters that could serve as drivers of such variations.  
Variations in the spectral index exhibit a pattern that is quite similar to $B_x-B_y$, which serves as an indicator of magnetic sector structure.
In particular, $L$ and $B_x-B_y$ exhibit a dominant period at $\approx$27 days, corresponding to solar rotation, which is much stronger near sunspot maximum than near sunspot minimum.
This implies that during this 8.5-year time interval, the solar wind predominantly had a two-sector structure, rather than the four-sector structure \citep[as in a classic figure by][]{Wilcox67} that would have a 13.5-day periodicity.
The GCR flux also exhibits a pattern that is most similar to that of $B_x-B_y$, but with behavior during a specific time period that indicates an additional influence from the solar wind speed.
In contrast, the magnetic field magnitude $B$ has a moderately dominant periodicity at 13.5 days, which is consistent with the two-sector structure, as $B$ should approach zero at sector boundaries, which occur twice per solar rotation.
This pattern is inconsistent with those of $C$ and $L$, which argues against the magnetic field magnitude as a source of their 27-day variations.
Because the cosmic ray diffusion coefficient is generally considered to vary inversely with $B$, this argues against a dominant role for a diffusive process in the 27-day modulation of cosmic rays at $\sim$10 GV.

The wavelet power pattern of the solar wind speed is also quite distinct from the pattern of the SP NM leader fraction.  
The solar wind speed varies with either a dominant 27-day period or a dominant 13.5-day period, alternating at different time periods, and maintains significant periodicity sporadically throughout the sunspot minimum period.  
Physically, a high speed solar wind stream at Earth usually results from a near-equatorial coronal hole or an equatorward extension of a polar coronal hole.
The presence of one coronal hole near the solar equator, in a specific range of heliolongitude, can lead to a 27-day period, whereas the presence of two such coronal holes at nearly opposite longitudes leads to a dominant 13.5-day period.  
The alternating (non-simultaneous) nature of the solar wind speed wavelet power observed at 27 days or 13.5 days supports this interpretation.
However, the spectral index \add{as} indicated by $L$ does not exhibit such alternation, but retains a dominant 27-day power (as does $B_x-B_y$) throughout these events, such as the time period in late 2016 that was discussed in detail in Section \ref{sec:Periodic_L}.
To the extent that the solar wind speed influences the cosmic ray spectrum, such influences are apparently nearly rigidity-independent, perhaps substantially affecting the cosmic ray flux (e.g., during late 2016 as described above) but not its spectral index.  
The possible influences of the solar wind speed periodicity -- which apparently do not affect the GCR spectral index at $\sim$10 GV -- include solar wind convection of scattering centers and interactions between faster solar wind streams and preceding slower streams (including CIRs). 

This analysis implies that 27-day periodicity in $L$ and the spectral index relates to the magnetic sector structure, which was predominantly a two-sector structure.  
A possible physical mechanism for such an influence would be drifts along the HCS at the sector boundary.
The direction of the drift - i.e., outward to inhibit the access of GCRs to the inner heliosphere or inward to enhance such access - depends on the polarity of the preceding and following magnetic sectors.  
Indeed, in a two-sector structure, the radial drift speed would positive during one HCS crossing and negative during the other, resulting in a 27-day periodicity.
In contrast, as noted above, a diffusive effect would respond to the magnetic field magnitude $B$, which exhibits a 13.5-day period for the same two-sector structure.
We therefore consider HCS drift to be a prime candidate to fully explain variations in the GCR spectral index at rigidity $\sim$10 GV.  
HCS drift can also largely explain the flux variations in this rigidity range, though it seems that at times they are substantially affected by the periodic variation of the solar wind speed, presumably associated with coronal holes in the solar equatorial region, which can influence GCRs by means of convection or CIRs.

Finally, we note other applications of the leader fraction from polar NMs that can be expected in the near future.
Following this demonstration of its utility, our team intends to make leader fraction data available in real time for monitoring of the cosmic ray spectral index.
Variations in the GCR spectral index are small, and in our experience daily averaging of $L$ from a polar NM is required for useful analysis, so for studies of GCR spectral variations ``real-time'' really implies time scales on the order of a day.
However, relativistic solar particles during a GLE event have a dramatically different spectral index, and their flux can evolve over time scales of minutes, possibly leading to changes in $L$ over shorter time scales that exceed statistical fluctuations.
Such a distinctive signal during a GLE event implies that monitoring of $L$ data from polar NM stations at hourly or finer time cadence could contribute to GLE alert systems, a possibility that we will investigate in further work.

\begin{center}
Acknowledgments
\end{center}

We gratefully acknowledge the logistical support provided by Australia’s Antarctic Program for operating the Mawson neutron monitor. 
This research was also supported by the National Science and Technology Development Agency
(NSTDA) and the National Research Council of Thailand (NRCT) under the High-Potential Research
Team Grant Program (N42A650868), by NRCT (N42A661044), by the NSRF via the Program
Management Unit for Human Resources \& Institutional Development, Research and Innovation
(B37G660015), and by the US NSF Award for Collaborative Research: The Simpson Neutron
Monitor Network (2112437) and predecessor awards for the South Pole Neutron
Monitor operation.
Suyeon Oh was supported by a National Research Foundation of Korea (NRF) grant funded by the Republic of Korea government (MSIT) (2023R1A2C1004304).

\bibliography{refff}{}

\begin{thebibliography}{}
\expandafter\ifx\csname natexlab\endcsname\relax\def\natexlab#1{#1}\fi
\providecommand{\url}[1]{\href{#1}{#1}}
\providecommand{\dodoi}[1]{doi:~\href{http://doi.org/#1}{\nolinkurl{#1}}}
\providecommand{\doeprint}[1]{\href{http://ascl.net/#1}{\nolinkurl{http://ascl.net/#1}}}
\providecommand{\doarXiv}[1]{\href{https://arxiv.org/abs/#1}{\nolinkurl{https://arxiv.org/abs/#1}}}

\bibitem[{{Abe} {et~al.}(2016){Abe}, {Fuke}, {Haino}, {Hams}, {Hasegawa}, {Horikoshi}, {Itazaki}, {Kim}, {Kumazawa}, {Kusumoto}, {Lee}, {Makida}, {Matsuda}, {Matsukawa}, {Matsumoto}, {Mitchell}, {Myers}, {Nishimura}, {Nozaki}, {Orito}, {Ormes}, {Picot-Clemente}, {Sakai}, {Sasaki}, {Seo}, {Shikaze}, {Shinoda}, {Streitmatter}, {Suzuki}, {Takasugi}, {Takeuchi}, {Tanaka}, {Thakur}, {Yamagami}, {Yamamoto}, {Yoshida}, \& {Yoshimura}}]{Abe16}
{Abe}, K., {Fuke}, H., {Haino}, S., {et~al.} 2016, \apj, 822, 65, \dodoi{10.3847/0004-637X/822/2/65}

\bibitem[{{Adriani} {et~al.}(2016){Adriani}, {Barbarino}, {Bazilevskaya}, {Bellotti}, {Boezio}, {Bogomolov}, {Bongi}, {Bonvicini}, {Bottai}, {Bruno}, {Cafagna}, {Campana}, {Carlson}, {Casolino}, {Castellini}, {De Santis}, {Di Felice}, {Galper}, {Karelin}, {Koldashov}, {Koldobskiy}, {Krutkov}, {Kvashnin}, {Leonov}, {Malakhov}, {Marcelli}, {Martucci}, {Mayorov}, {Menn}, {Merg{\'e}}, {Mikhailov}, {Mocchiutti}, {Monaco}, {Mori}, {Munini}, {Osteria}, {Panico}, {Papini}, {Pearce}, {Picozza}, {Ricci}, {Ricciarini}, {Simon}, {Sparvoli}, {Spillantini}, {Stozhkov}, {Vacchi}, {Vannuccini}, {Vasilyev}, {Voronov}, {Yurkin}, {Zampa}, {Zampa}, {Potgieter}, \& {Vos}}]{Adriani16}
{Adriani}, O., {Barbarino}, G.~C., {Bazilevskaya}, G.~A., {et~al.} 2016, \prl, 116, 241105, \dodoi{10.1103/PhysRevLett.116.241105}

\bibitem[{{Aguilar} {et~al.}(2021){Aguilar}, {Cavasonza}, {Ambrosi}, {Arruda}, {Attig}, {Barao}, {Barrin}, {Bartoloni}, {Ba{\c{s}}e{\v{g}}mez-du Pree}, {Battiston}, {Behlmann}, {Beranek}, {Berdugo}, {Bertucci}, {Bindi}, {Bollweg}, {Borgia}, {Boschini}, {Bourquin}, {Bueno}, {Burger}, {Burger}, {Burmeister}, {Cai}, {Capell}, {Casaus}, {Castellini}, {Cervelli}, {Chang}, {Chen}, {Chen}, {Chen}, {Chen}, {Cheng}, {Chou}, {Chouridou}, {Choutko}, {Chung}, {Clark}, {Coignet}, {Consolandi}, {Contin}, {Corti}, {Cui}, {Dadzie}, {Dass}, {Delgado}, {Della Torre}, {Demirk{\"o}z}, {Derome}, {Di Falco}, {Di Felice}, {D{\'\i}az}, {Dimiccoli}, {von Doetinchem}, {Dong}, {Donnini}, {Duranti}, {Egorov}, {Eline}, {Feng}, {Fiandrini}, {Fisher}, {Formato}, {Freeman}, {G{\'a}mez}, {Garc{\'\i}a-L{\'o}pez}, {Gargiulo}, {Gast}, {Gervasi}, {Giovacchini}, {G{\'o}mez-Coral}, {Gong}, {Goy}, {Grabski}, {Grandi}, {Graziani}, {Haino}, {Han}, {Hashmani}, {He}, {Heber}, {Hsieh}, {Hu}, {Incagli}, {Jang}, {Jia}, {Jinchi}, {Karag{\"o}z}, {Khiali},
  {Kim}, {Kirn}, {Konyushikhin}, {Kounina}, {Kounine}, {Koutsenko}, {Krasnopevtsev}, {Kuhlman}, {Kulemzin}, {La Vacca}, {Laudi}, {Laurenti}, {Lazzizzera}, {Lebedev}, {Lee}, {Lee}, {Li}, {Li}, {Li}, {Li}, {Li}, {Li}, {Liang}, {Light}, {Lin}, {Lippert}, {Liu}, {Liu}, {Lu}, {Lu}, {Luebelsmeyer}, {Luo}, {Luo}, {Machate}, {Ma{\~n}{\'a}}, {Mar{\'\i}n}, {Marquardt}, {Martin}, {Mart{\'\i}nez}, {Masi}, {Maurin}, {Medvedeva}, {Menchaca-Rocha}, {Meng}, {Mikhailov}, {Molero}, {Mott}, {Mussolin}, {Negrete}, {Nikonov}, {Nozzoli}, {Oliva}, {Orcinha}, {Palermo}, {Palmonari}, {Paniccia}, {Pashnin}, {Pauluzzi}, {Pensotti}, {Phan}, {Plyaskin}, {Pohl}, {Poluianov}, {Qin}, {Qu}, {Quadrani}, {Rancoita}, {Rapin}, {Conde}, {Robyn}, {Rosier-Lees}, {Rozhkov}, {Rozza}, {Sagdeev}, {Schael}, {von Dratzig}, {Schwering}, {Seo}, {Shakfa}, {Shan}, {Siedenburg}, {Solano}, {Song}, {Song}, {Sonnabend}, {Strigari}, {Su}, {Sun}, {Sun}, {Tacconi}, {Tang}, {Tang}, {Tian}, {Ting}, {Ting}, {Tomassetti}, {Torsti}, {Urban}, {Usoskin}, {Vagelli},
  {Vainio}, {Valencia-Otero}, {Valente}, {Valtonen}, {V{\'a}zquez Acosta}, {Vecchi}, {Velasco}, {Vialle}, {Wang}, {Wang}, {Wang}, {Wang}, {Wang}, {Wang}, {Wang}, {Wang}, {Wang}, {Wei}, {Weng}, {Wu}, {Xiong}, {Xu}, {Yan}, {Yang}, {Yashin}, {Yi}, {Yu}, {Yu}, {Zannoni}, {Zhang}, {Zhang}, {Zhang}, {Zhang}, {Zhang}, {Zhao}, {Zheng}, {Zheng}, {Zhuang}, {Zhukov}, {Zichichi}, {Zuccon}, \& {AMS Collaboration}}]{Aguilar21}
{Aguilar}, M., {Cavasonza}, L.~A., {Ambrosi}, G., {et~al.} 2021, \prl, 127, 271102, \dodoi{10.1103/PhysRevLett.127.271102}

\bibitem[{{Aguilar} {et~al.}(2022){Aguilar}, {Cavasonza}, {Ambrosi}, {Arruda}, {Attig}, {Barao}, {Barrin}, {Bartoloni}, {Ba{\c{s}}e{\v{g}}mez-du Pree}, {Battiston}, {Behlmann}, {Berdugo}, {Bertucci}, {Bindi}, {Bollweg}, {Borgia}, {Boschini}, {Bourquin}, {Bueno}, {Burger}, {Burger}, {Burmeister}, {Cai}, {Capell}, {Casaus}, {Castellini}, {Cervelli}, {Chang}, {Chen}, {Chen}, {Chen}, {Chen}, {Cheng}, {Chou}, {Chouridou}, {Choutko}, {Chung}, {Clark}, {Coignet}, {Consolandi}, {Contin}, {Corti}, {Cui}, {Dadzie}, {Dass}, {Delgado}, {Della Torre}, {Demirk{\"o}z}, {Derome}, {Di Falco}, {Di Felice}, {D{\'\i}az}, {Dimiccoli}, {von Doetinchem}, {Dong}, {Donnini}, {Duranti}, {Egorov}, {Eline}, {Feng}, {Fiandrini}, {Fisher}, {Formato}, {Freeman}, {G{\'a}mez}, {Garc{\'\i}a-L{\'o}pez}, {Gargiulo}, {Gast}, {Gervasi}, {Giovacchini}, {G{\'o}mez-Coral}, {Gong}, {Goy}, {Grabski}, {Grandi}, {Graziani}, {Haino}, {Han}, {Hashmani}, {He}, {Heber}, {Hsieh}, {Hu}, {Incagli}, {Jang}, {Jia}, {Jinchi}, {Karag{\"o}z}, {Khiali}, {Kim},
  {Kirn}, {Konyushikhin}, {Kounina}, {Kounine}, {Koutsenko}, {Krasnopevtsev}, {Kuhlman}, {Kulemzin}, {La Vacca}, {Laudi}, {Laurenti}, {Lazzizzera}, {Lee}, {Lee}, {Li}, {Li}, {Li}, {Li}, {Li}, {Li}, {Li}, {Li}, {Li}, {Liang}, {Liang}, {Light}, {Lin}, {Lippert}, {Liu}, {Lu}, {Lu}, {Luebelsmeyer}, {Luo}, {Luo}, {Machate}, {Ma{\~n}{\'a}}, {Mar{\'\i}n}, {Marquardt}, {Martin}, {Mart{\'\i}nez}, {Masi}, {Maurin}, {Medvedeva}, {Menchaca-Rocha}, {Meng}, {Mikhailov}, {Molero}, {Mott}, {Mussolin}, {Negrete}, {Nikonov}, {Nozzoli}, {Ocampo-Peleteiro}, {Oliva}, {Orcinha}, {Palermo}, {Palmonari}, {Paniccia}, {Pashnin}, {Pauluzzi}, {Pensotti}, {Plyaskin}, {Pohl}, {Poluianov}, {Qin}, {Qu}, {Quadrani}, {Rancoita}, {Rapin}, {Conde}, {Robyn}, {Rosier-Lees}, {Rozhkov}, {Rozza}, {Sagdeev}, {Schael}, {von Dratzig}, {Schwering}, {Seo}, {Shan}, {Siedenburg}, {Song}, {Song}, {Sonnabend}, {Strigari}, {Su}, {Sun}, {Sun}, {Tacconi}, {Tang}, {Tang}, {Tian}, {Ting}, {Ting}, {Tomassetti}, {Torsti}, {Urban}, {Usoskin}, {Vagelli}, {Vainio},
  {Valencia-Otero}, {Valente}, {Valtonen}, {V{\'a}zquez Acosta}, {Vecchi}, {Velasco}, {Vialle}, {Wang}, {Wang}, {Wang}, {Wang}, {Wang}, {Wang}, {Wang}, {Wang}, {Wang}, {Wei}, {Weng}, {Wu}, {Xiong}, {Xu}, {Yan}, {Yang}, {Yashin}, {Yi}, {Yu}, {Yu}, {Zannoni}, {Zhang}, {Zhang}, {Zhang}, {Zhang}, {Zhang}, {Zhao}, {Zheng}, {Zheng}, {Zhuang}, {Zhukov}, {Zichichi}, {Zuccon}, \& {AMS Collaboration}}]{Aguilar22}
---. 2022, \prl, 128, 231102, \dodoi{10.1103/PhysRevLett.128.231102}

\bibitem[{{Aguilar} {et~al.}(2023){Aguilar}, {Ambrosi}, {Anderson}, {Arruda}, {Attig}, {Bagwell}, {Barao}, {Barbanera}, {Barrin}, {Bartoloni}, {Battiston}, {Belyaev}, {Berdugo}, {Bertucci}, {Bindi}, {Bollweg}, {Bolster}, {Borchiellini}, {Borgia}, {Boschini}, {Bourquin}, {Burger}, {Burger}, {Cai}, {Capell}, {Casaus}, {Castellini}, {Cervelli}, {Chang}, {Chen}, {Chen}, {Chen}, {Chen}, {Chen}, {Cheng}, {Chou}, {Chouridou}, {Choutko}, {Chung}, {Clark}, {Coignet}, {Consolandi}, {Contin}, {Corti}, {Cui}, {Dadzie}, {D'Angelo}, {Dass}, {Delgado}, {Della Torre}, {Demirk{\"o}z}, {Derome}, {Di Falco}, {Di Felice}, {D{\'\i}az}, {Dimiccoli}, {von Doetinchem}, {Dong}, {Donnini}, {Duranti}, {Egorov}, {Eline}, {Faldi}, {Feng}, {Fiandrini}, {Fisher}, {Formato}, {G{\'a}mez}, {Garc{\'\i}a-L{\'o}pez}, {Gargiulo}, {Gast}, {Gervasi}, {Giovacchini}, {G{\'o}mez-Coral}, {Gong}, {Goy}, {Grandi}, {Graziani}, {Guracho}, {Haino}, {Han}, {Hashmani}, {He}, {Heber}, {Hsieh}, {Hu}, {Huang}, {Ionica}, {Incagli}, {Jia}, {Jinchi}, {Karag{\"o}z},
  {Khan}, {Khiali}, {Kirn}, {Klipfel}, {Kounina}, {Kounine}, {Koutsenko}, {Krasnopevtsev}, {Kuhlman}, {Kulemzin}, {La Vacca}, {Laudi}, {Laurenti}, {LaVecchia}, {Lazzizzera}, {Lee}, {Lee}, {Li}, {Li}, {Li}, {Li}, {Li}, {Li}, {Li}, {Li}, {Li}, {Li}, {Li}, {Liang}, {Liang}, {Lin}, {Lippert}, {Liu}, {Lu}, {Lu}, {Luebelsmeyer}, {Luo}, {Luo}, {Luo}, {Ma{\~n}{\'a}}, {Mar{\'\i}n}, {Marquardt}, {Martin}, {Mart{\'\i}nez}, {Masi}, {Maurin}, {Medvedeva}, {Menchaca-Rocha}, {Meng}, {Molero}, {Mott}, {Mussolin}, {Jozani}, {Negrete}, {Nicolaidis}, {Nikonov}, {Nozzoli}, {Ocampo-Peleteiro}, {Oliva}, {Orcinha}, {Ottupara}, {Palermo}, {Palmonari}, {Paniccia}, {Pashnin}, {Pauluzzi}, {Pensotti}, {Plyaskin}, {Poluianov}, {Qin}, {Qu}, {Quadrani}, {Rancoita}, {Rapin}, {Conde}, {Robyn}, {Rodr{\'\i}guez-Garc{\'\i}a}, {Romaneehsen}, {Rossi}, {Rozhkov}, {Rozza}, {Sagdeev}, {Savin}, {Schael}, {von Dratzig}, {Schwering}, {Seo}, {Shan}, {Siedenburg}, {Silvestre}, {Song}, {Song}, {Sonnabend}, {Strigari}, {Su}, {Sun}, {Sun}, {Tacconi},
  {Tang}, {Tang}, {Tian}, {Tian}, {Ting}, {Ting}, {Tomassetti}, {Torsti}, {Urban}, {Usoskin}, {Vagelli}, {Vainio}, {Valencia-Otero}, {Valente}, {Valtonen}, {V{\'a}zquez Acosta}, {Vecchi}, {Velasco}, {Vialle}, {Wang}, {Wang}, {Wang}, {Wang}, {Wang}, {Wang}, {Wang}, {Wang}, {Wang}, {Wei}, {Weng}, {Wu}, {Wu}, {Xiao}, {Xiong}, {Xiong}, {Xu}, {Yan}, {Yang}, {Yang}, {Yelland}, {Yi}, {You}, {Yu}, {Yu}, {Zhang}, {Zhang}, {Zhang}, {Zhang}, {Zhang}, {Zhang}, {Zhao}, {Zheng}, {Zheng}, {Zhuang}, {Zhukov}, {Zichichi}, {Zuccon}, \& {AMS Collaboration}}]{Aguilar23}
{Aguilar}, M., {Ambrosi}, G., {Anderson}, H., {et~al.} 2023, \prl, 131, 151002, \dodoi{10.1103/PhysRevLett.131.151002}

\bibitem[{{Banglieng} {et~al.}(2020){Banglieng}, {Janthaloet}, {Ruffolo}, {S{\'a}iz}, {Mitthumsiri}, {Muangha}, {Evenson}, {Nutaro}, {Pyle}, {Seunarine}, {Madsen}, {Mangeard}, \& {Macatangay}}]{Banglieng20}
{Banglieng}, C., {Janthaloet}, H., {Ruffolo}, D., {et~al.} 2020, \apj, 890, 21, \dodoi{10.3847/1538-4357/ab6661}

\bibitem[{{Bieber} {et~al.}(2004){Bieber}, {Clem}, {Duldig}, {Evenson}, {Humble}, \& {Pyle}}]{BieberEA04}
{Bieber}, J.~W., {Clem}, J.~M., {Duldig}, M.~L., {et~al.} 2004, J. Geophys. Res. (Space Phys.), 109, A12106, \dodoi{10.1029/2004JA010493}

\bibitem[{{Buatthaisong} {et~al.}(2022){Buatthaisong}, {Ruffolo}, {S{\'a}iz}, {Banglieng}, {Mitthumsiri}, {Nutaro}, \& {Nuntiyakul}}]{Buatthaisong22}
{Buatthaisong}, N., {Ruffolo}, D., {S{\'a}iz}, A., {et~al.} 2022, \apj, 939, 99, \dodoi{10.3847/1538-4357/ac96ea}

\bibitem[{{Clem} \& {Dorman}(2000)}]{Clem00}
{Clem}, J.~M., \& {Dorman}, L.~I. 2000, \ssr, 93, 335, \dodoi{10.1023/A:1026508915269}

\bibitem[{{Dr{\"o}ge} {et~al.}(1990){Dr{\"o}ge}, {Wibberenz}, \& {Klecker}}]{Droege90}
{Dr{\"o}ge}, W., {Wibberenz}, G., \& {Klecker}, B. 1990, in International Cosmic Ray Conference, Vol.~5, 21st International Cosmic Ray Conference (ICRC21), 187

\bibitem[{{Engelbrecht} {et~al.}(2022){Engelbrecht}, {Effenberger}, {Florinski}, {Potgieter}, {Ruffolo}, {Chhiber}, {Usmanov}, {Rankin}, \& {Els}}]{Engelbrecht22}
{Engelbrecht}, N.~E., {Effenberger}, F., {Florinski}, V., {et~al.} 2022, \ssr, 218, 33, \dodoi{10.1007/s11214-022-00896-1}

\bibitem[{{Evenson} {et~al.}(2022){Evenson}, {Clem}, {Mangeard}, {Nuntiyakul}, {Ruffolo}, {S{\'a}iz}, {Seripienlert}, \& {Seunarine}}]{Evenson22}
{Evenson}, P., {Clem}, J., {Mangeard}, P.~S., {et~al.} 2022, in 37th International Cosmic Ray Conference, 1240, \dodoi{10.22323/1.395.01240}

\bibitem[{{Fonger}(1953)}]{Fonger53}
{Fonger}, W.~H. 1953, Physical Review, 91, 351, \dodoi{10.1103/PhysRev.91.351}

\bibitem[{{Forbush}(1937)}]{Forbush37}
{Forbush}, S.~E. 1937, Physical Review, 51, 1108, \dodoi{10.1103/PhysRev.51.1108.3}

\bibitem[{{Forbush}(1954)}]{Forbush54}
---. 1954, \jgr, 59, 525, \dodoi{10.1029/JZ059i004p00525}

\bibitem[{{Garcia-Munoz} {et~al.}(1986){Garcia-Munoz}, {Meyer}, {Pyle}, {Simpson}, \& {Evenson}}]{GarciaMunoz86}
{Garcia-Munoz}, M., {Meyer}, P., {Pyle}, K.~R., {Simpson}, J.~A., \& {Evenson}, P. 1986, \jgr, 91, 2858, \dodoi{10.1029/JA091iA03p02858}

\bibitem[{{Ghanbari} {et~al.}(2019){Ghanbari}, {Florinski}, {Guo}, {Hu}, \& {Leske}}]{Ghanbari19}
{Ghanbari}, K., {Florinski}, V., {Guo}, X., {Hu}, Q., \& {Leske}, R. 2019, \apj, 882, 54, \dodoi{10.3847/1538-4357/ab31a5}

\bibitem[{{Ghelfi} {et~al.}(2016){Ghelfi}, {Barao}, {Derome}, \& {Maurin}}]{Ghelfi16}
{Ghelfi}, A., {Barao}, F., {Derome}, L., \& {Maurin}, D. 2016, \aap, 591, A94, \dodoi{10.1051/0004-6361/201527852}

\bibitem[{{Hatton} \& {Carmichael}(1964)}]{Hatton64}
{Hatton}, C.~J., \& {Carmichael}, H. 1964, Canadian Journal of Physics, 42, 2443, \dodoi{10.1139/p64-222}

\bibitem[{{Hess} \& {Graziadei}(1936)}]{Hess36}
{Hess}, V.~F., \& {Graziadei}, H.~T. 1936, Terrestrial Magnetism and Atmospheric Electricity, 41, 9, \dodoi{10.1029/TE041i001p00009}

\bibitem[{{Jokipii} \& {Thomas}(1981)}]{Jokipii81}
{Jokipii}, J.~R., \& {Thomas}, B. 1981, \apj, 243, 1115, \dodoi{10.1086/158675}

\bibitem[{{Jung} {et~al.}(2016){Jung}, {Oh}, {Yi}, {Evenson}, {Pyle}, {Jee}, {Kim}, {Lee}, \& {Sohn}}]{Jung16}
{Jung}, J., {Oh}, S., {Yi}, Y., {et~al.} 2016, Journal of Astronomy and Space Sciences, 33, 345, \dodoi{10.5140/JASS.2016.33.4.345}

\bibitem[{Kittiya {et~al.}(2023)Kittiya, Nuntiyakul, Seripienlert, Saiz~Rivera, Ruffolo, Evenson, Sonsrettee, \& Oh}]{Kittiy23}
Kittiya, A., Nuntiyakul, W., Seripienlert, A., {et~al.} 2023, in 38th International Cosmic Ray Conference, 1320, \dodoi{10.22323/1.444.1320}

\bibitem[{{Mangeard} {et~al.}(2016{\natexlab{a}}){Mangeard}, {Ruffolo}, {S{\'a}iz}, {Madlee}, \& {Nutaro}}]{Mangeard16-PSNM}
{Mangeard}, P.~S., {Ruffolo}, D., {S{\'a}iz}, A., {Madlee}, S., \& {Nutaro}, T. 2016{\natexlab{a}}, J. Geophys. Res. (Space Phys.), 121, 7435, \dodoi{10.1002/2016JA022638}

\bibitem[{{Mangeard} {et~al.}(2016{\natexlab{b}}){Mangeard}, {Ruffolo}, {S{\'a}iz}, {Nuntiyakul}, {Bieber}, {Clem}, {Evenson}, {Pyle}, {Duldig}, \& {Humble}}]{Mangeard16-latsur}
{Mangeard}, P.~S., {Ruffolo}, D., {S{\'a}iz}, A., {et~al.} 2016{\natexlab{b}}, Journal of Geophysical Research (Space Physics), 121, 11,620, \dodoi{10.1002/2016JA023515}

\bibitem[{{Modzelewska} \& {Gil}(2021)}]{Modzelewska21a}
{Modzelewska}, R., \& {Gil}, A. 2021, \aap, 646, A128, \dodoi{10.1051/0004-6361/202039651}

\bibitem[{{Modzelewska} {et~al.}(2019){Modzelewska}, {Mayorov}, {Munini}, \& {PAMELA Collaboration}}]{Modzelewska19}
{Modzelewska}, R., {Mayorov}, A., {Munini}, R., \& {PAMELA Collaboration}. 2019, in Journal of Physics Conference Series, Vol. 1181, Journal of Physics Conference Series, 012015, \dodoi{10.1088/1742-6596/1181/1/012015}

\bibitem[{{Moraal} {et~al.}(1989){Moraal}, {Potgieter}, {Stoker}, \& {van der Walt}}]{Moraal89}
{Moraal}, H., {Potgieter}, M.~S., {Stoker}, P.~H., \& {van der Walt}, A.~J. 1989, \jgr, 94, 1459, \dodoi{10.1029/JA094iA02p01459}

\bibitem[{{Moses}(1987)}]{Moses87}
{Moses}, D. 1987, \apj, 313, 471, \dodoi{10.1086/164987}

\bibitem[{{Muangha} {et~al.}(2023){Muangha}, {Ruffolo}, {S{\'a}iz}, {Banglieng}, {Evenson}, \& {Seunarine}}]{muangha23}
{Muangha}, P., {Ruffolo}, D., {S{\'a}iz}, A., {et~al.} 2023, in 38th International Cosmic Ray Conference, 1304, \dodoi{10.22323/1.444.1304}

\bibitem[{{Muangha} {et~al.}(2022){Muangha}, {Ruffolo}, {S{\'a}iz}, {Banglieng}, {Evenson}, {Seunarine}, {Oh}, {Jung}, {Duldig}, \& {Humble}}]{Muangha21}
{Muangha}, P., {Ruffolo}, D., {S{\'a}iz}, A., {et~al.} 2022, in 37th International Cosmic Ray Conference, 1285, \dodoi{10.22323/1.395.01285}

\bibitem[{{Munakata} {et~al.}(2022){Munakata}, {Kozai}, {Kato}, {Hayashi}, {Kataoka}, {Kadokura}, {Tokumaru}, {Mendon{\c{c}}a}, {Echer}, {Lago}, {Rockenbach}, {Schuch}, {Bageston}, {Braga}, {Jassar}, {Sharma}, {Duldig}, {Humble}, {Sabbah}, {Evenson}, {Mangeard}, {Kuwabara}, {Ruffolo}, {S{\'a}iz}, {Mitthumsiri}, {Nuntiyakul}, \& {K{\'o}ta}}]{Munakata22}
{Munakata}, K., {Kozai}, M., {Kato}, C., {et~al.} 2022, \apj, 938, 30, \dodoi{10.3847/1538-4357/ac91c5}

\bibitem[{{Nuntiyakul} {et~al.}(2014){Nuntiyakul}, {Evenson}, {Ruffolo}, {S{\'a}iz}, {Bieber}, {Clem}, {Pyle}, {Duldig}, \& {Humble}}]{Nuntiyakul14}
{Nuntiyakul}, W., {Evenson}, P., {Ruffolo}, D., {et~al.} 2014, \apj, 795, 11, \dodoi{10.1088/0004-637X/795/1/11}

\bibitem[{{Parker}(1958)}]{Parker58}
{Parker}, E.~N. 1958, \apj, 128, 664, \dodoi{10.1086/146579}

\bibitem[{{Poluianov} {et~al.}(2017){Poluianov}, {Usoskin}, {Mishev}, {Shea}, \& {Smart}}]{Poluianov17}
{Poluianov}, S.~V., {Usoskin}, I.~G., {Mishev}, A.~L., {Shea}, M.~A., \& {Smart}, D.~F. 2017, \solphys, 292, 176, \dodoi{10.1007/s11207-017-1202-4}

\bibitem[{{Poopakun} {et~al.}(2023){Poopakun}, {Nuntiyakul}, {Khamphakdee}, {Seripienlert}, {Ruffolo}, {Evenson}, {Jiang}, {Chuanraksasat}, {Munakata}, {Duldig}, {Humble}, {Madsen}, {Soonthornthum}, \& {Komonjinda}}]{Poopakun23}
{Poopakun}, K., {Nuntiyakul}, W., {Khamphakdee}, S., {et~al.} 2023, \apj, 958, 80, \dodoi{10.3847/1538-4357/ad02f1}

\bibitem[{{Potgieter}(2013)}]{Potgieter13}
{Potgieter}, M.~S. 2013, Living Reviews in Solar Physics, 10, 3, \dodoi{10.12942/lrsp-2013-3}

\bibitem[{{Ruffolo}(2020)}]{Ruffolo20}
{Ruffolo}, D. 2020, in Journal of Physics Conference Series, Vol. 1572, Journal of Physics Conference Series, 012087, \dodoi{10.1088/1742-6596/1572/1/012087}

\bibitem[{{Ruffolo} {et~al.}(2016){Ruffolo}, {S{\'a}iz}, {Mangeard}, {Kamyan}, {Muangha}, {Nutaro}, {Sumran}, {Chaiwattana}, {Gasiprong}, {Channok}, {Wuttiya}, {Rujiwarodom}, {Tooprakai}, {Asavapibhop}, {Bieber}, {Clem}, {Evenson}, \& {Munakata}}]{Ruffolo16}
{Ruffolo}, D., {S{\'a}iz}, A., {Mangeard}, P.~S., {et~al.} 2016, \apj, 817, 38, \dodoi{10.3847/0004-637X/817/1/38}

\bibitem[{{S{\'a}iz} {et~al.}(2017){S{\'a}iz}, {Mitthumsiri}, {Ruffolo}, {Evenson}, \& {Nutaro}}]{Saiz17}
{S{\'a}iz}, A., {Mitthumsiri}, W., {Ruffolo}, D., {Evenson}, P., \& {Nutaro}, T. 2017, in International Cosmic Ray Conference, Vol. 301, 35th International Cosmic Ray Conference (ICRC2017), 47, \dodoi{10.22323/1.301.0047}

\bibitem[{{S{\'a}iz} {et~al.}(2019){S{\'a}iz}, {Mitthumsiri}, {Ruffolo}, {Evenson}, \& {Nutaro}}]{Saiz19}
{S{\'a}iz}, A., {Mitthumsiri}, W., {Ruffolo}, D., {Evenson}, P., \& {Nutaro}, T. 2019, in International Cosmic Ray Conference, Vol.~36, 36th International Cosmic Ray Conference (ICRC2019), 1145, \dodoi{10.22323/1.358.01145}

\bibitem[{{Simpson}(1948)}]{Simpson48}
{Simpson}, J.~A. 1948, Physical Review, 73, 1389, \dodoi{10.1103/PhysRev.73.1389}

\bibitem[{{Smart} \& {Shea}(2009)}]{Smart09}
{Smart}, D.~F., \& {Shea}, M.~A. 2009, Advances in Space Research, 44, 1107, \dodoi{10.1016/j.asr.2009.07.005}

\bibitem[{{Usoskin} {et~al.}(2005){Usoskin}, {Alanko-Huotari}, {Kovaltsov}, \& {Mursula}}]{Usoskin05}
{Usoskin}, I.~G., {Alanko-Huotari}, K., {Kovaltsov}, G.~A., \& {Mursula}, K. 2005, Journal of Geophysical Research (Space Physics), 110, A12108, \dodoi{10.1029/2005JA011250}

\bibitem[{{V{\"a}is{\"a}nen} {et~al.}(2023){V{\"a}is{\"a}nen}, {Usoskin}, {K{\"a}hk{\"o}nen}, {Koldobskiy}, \& {Mursula}}]{Vaisanen23}
{V{\"a}is{\"a}nen}, P., {Usoskin}, I., {K{\"a}hk{\"o}nen}, R., {Koldobskiy}, S., \& {Mursula}, K. 2023, Journal of Geophysical Research (Space Physics), 128, e2023JA031352, \dodoi{10.1029/2023JA031352}

\bibitem[{{Wilcox} \& {Ness}(1967)}]{Wilcox67}
{Wilcox}, J.~M., \& {Ness}, N.~F. 1967, \solphys, 1, 437, \dodoi{10.1007/BF00151368}

\end{thebibliography}

\appendix

\section{Normalization of the Leader Fraction}

\citet{Banglieng20} discussed the use of different series of electronics firmware; in the present work only data from 800 series firmware are analyzed.  
However, as part of the development of our data acquisition system, there have been changes in the version of the data acquisition software (called ``LandMonitor'') at the South Pole (SP), Jang Bogo (JB), and Mawson (MA) stations, as well as some physical changes to the JB NM configuration. 
Here we describe such changes and show the raw leader fraction data.
For subsequent analysis, the $L$ values were averaged over counter tubes (correcting for any tubes with missing data) and normalized to correct for changes in software or physical configuration for any version or configuration used longer than two months.  
They were normalized to the level of the latest software version and configuration, and were then corrected for variations in atmospheric pressure.


\begin{figure}
	\centering
        \includegraphics[scale = 0.42]{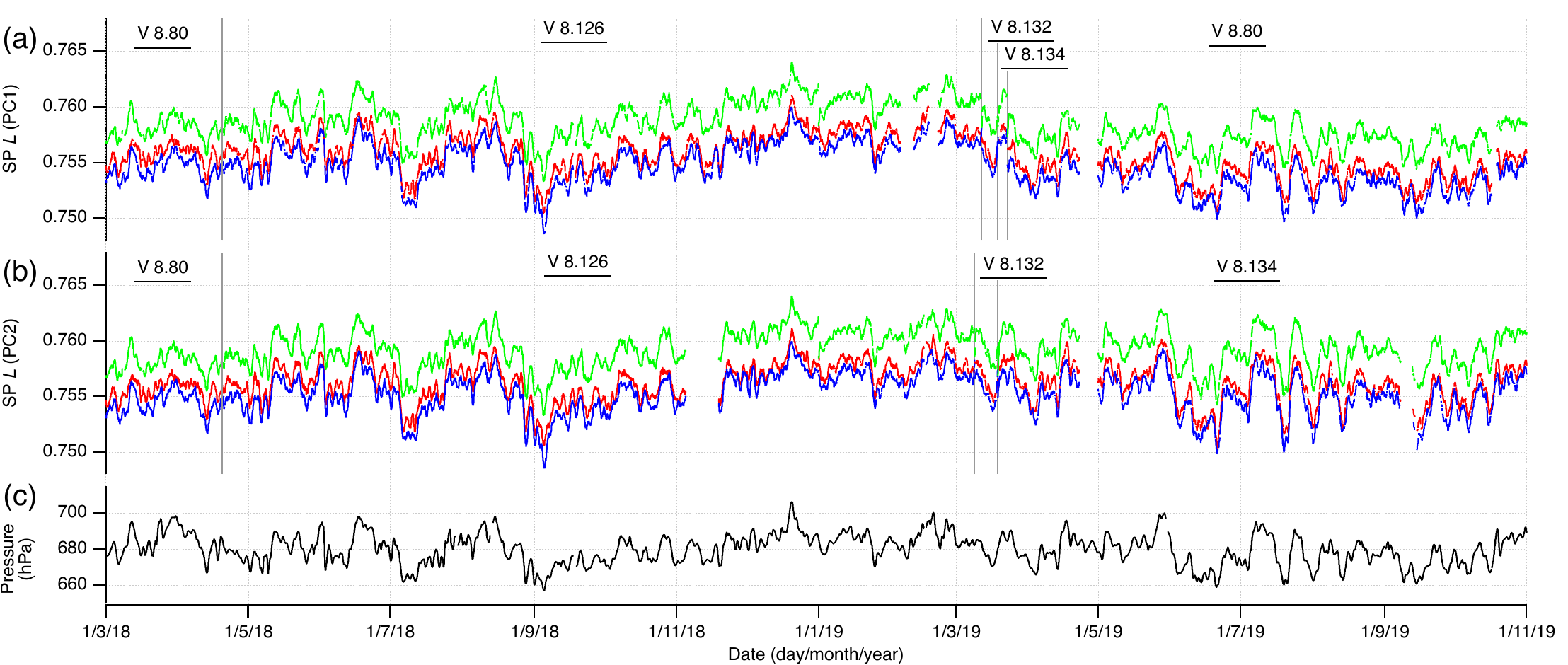}
\caption{ Daily raw leader fraction $L$ of South Pole (SP) NM during a time period with software changes, 2018 March through 2019 October. The daily $L$ values were extracted from (a) computer 1 (PC1) and (b)  computer 2 (PC2). $L$ values from each of 3 counter tubes are indicated by colored lines. (c) SP pressure in hPa. The SP NM has recorded time-delay histograms on two computers simultaneously.  Vertical lines indicate the times when data acquisition software was changed.}
	\label{fig.SP_cali}
\end{figure}
\subsection{South Pole NM station}

Figures \ref{fig.SP_cali}(a) and (b) present the raw SP $L$ for each NM counter from 2018 March through 2019 October.
As described in Section \ref{sec:Extract_L}, the leader fraction was determined from a fit to the exponential tail of each hourly time delay histogram from each counter tube. 
\citet{Banglieng20} previously normalized the SP $L$ from March 2015 to April 2018.
On 2018 April 20, the data acquisition software \add{``LandMonitor''} was switched from version 8.80 to 8.126. The fitting chi-squared ($\chi^2$) when extracting $L$ from the time-delay histograms was found to fluctuate strongly for LandMonitor 8.126.  
However, the value of $L$ was not affected, as we did not observe a significantly change when changing from 8.80 to 8.126.
Note that SP NM data were collected on two computers (PC1 and PC2) simultaneously.
On 2019 March 8, computer number 2 (PC2) started running LandMonitor 8.132. 
On 2019 March 11, PC1 was switched to 
\add{also} run LandMonitor 8.132 
\add{for our ``simultaneous test.''}
Both PC1 and PC2 at the South Pole were updated to run LandMonitor 8.134 on 2019 March 18.
\add{During the simultaneous test, we measured extremely small differences between the hourly values of $L$ between PC1 and PC2 when they operated the same software, with an average difference of $6.30\times10^{-7}$ and a standard deviation of $1.94\times10^{-5}$, perhaps associated with clock differences. For further analysis, these differences are considered negligible, and only data from PC1 are used.}
On 2019 March 22, PC1 was switched back to LandMonitor 8.80  with PC2 still running LandMonitor 8.134.
Since 2019 Sep 12, PC2 has produced 2 sets of histograms, one set like 8.80 and one set like 8.134 (the latter set is not shown in Figure \ref{fig.SP_cali}).

\begin{figure}
	\centering
	\includegraphics[scale = 0.42]{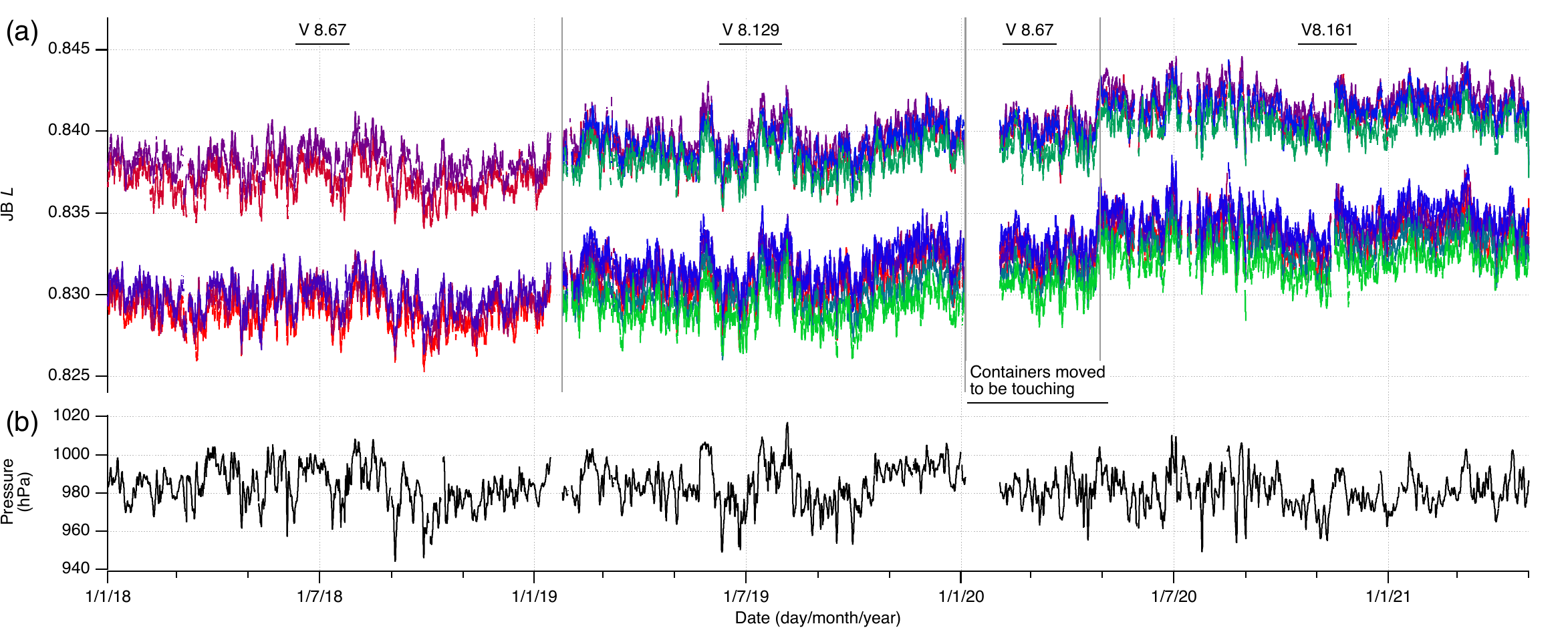}
	\caption{(a) 
 Daily raw leader fraction $L$ of Jang Bogo (JB) NM during a time period with changes in software and physical configuration, 2018 November through 2021 April, indicated by a colored trace for each counter tube.
(b) JB atmospheric pressure in hPa. 
 Vertical lines indicate the times of software or configuration changes. }
	\label{fig.JB_cali}
\end{figure}

\subsection{Jang Bogo NM station}
Raw JB $L$ values are shown in Figure \ref{fig.JB_cali} from 2018 January through 2021 April.  Note that the end tube in each row of three counter tubes has a lower value of $L$, as found in previous work \citep{Ruffolo16}.
From 2015 December to 2019 January, JB NM had one container with 6 tubes (in two rows of three tubes each, at the front and back of the container, separated by polyethylene reflector), including 5 NM tubes equipped with special electronics, and was running LandMonitor 8.67.
In 2019 January, two more containers of 6 tubes each were installed at Jang Bogo; \add{all} 3 container-loads of NM equipment came from McMurdo Station. From 2019 Jan 24 to 2020 Jan 5, JB NM operated using 8.129.
Initially physically separated, the containers were moved to be in contact with each other in 2020 January, and the LandMonitor version reverted back to 8.67. After 2020 April 29, JB NM operated using 8.161.

\begin{figure}
	\centering
	\includegraphics[scale = 0.42]{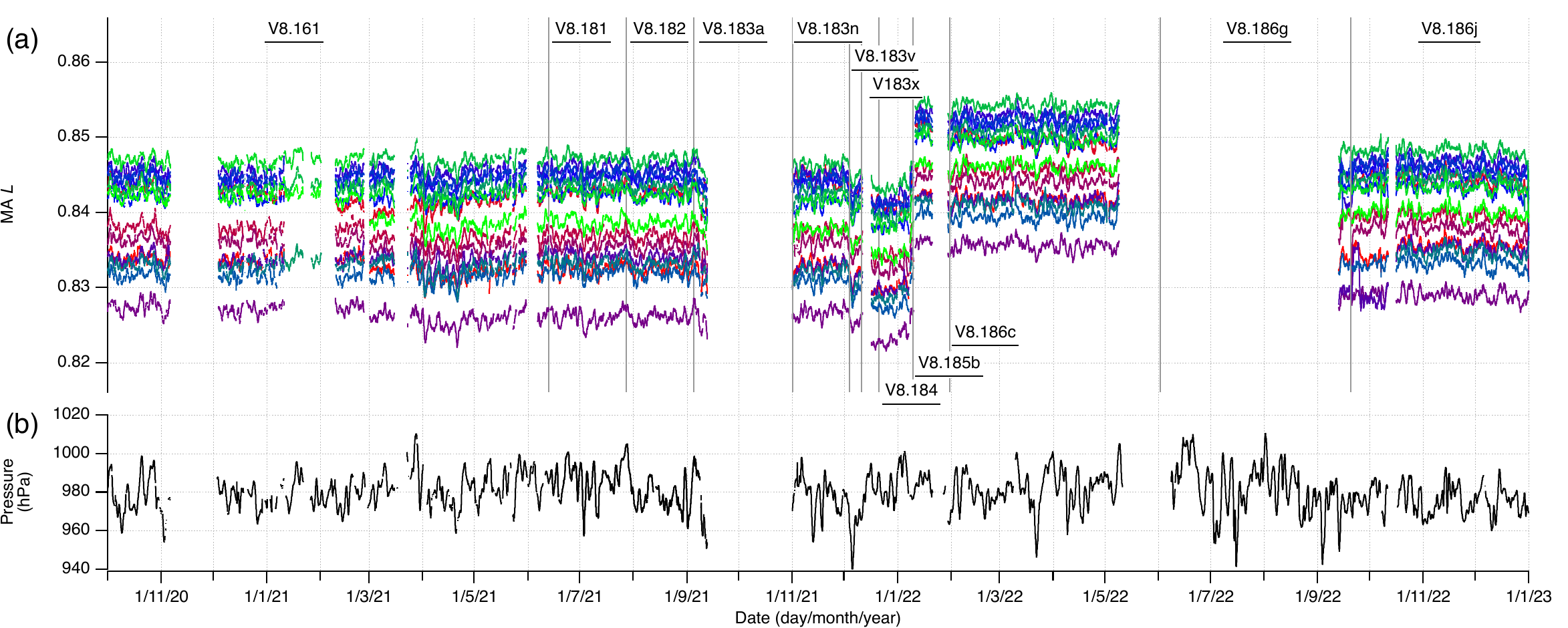}
	\caption{
(a) 
 Daily raw leader fraction $L$ of Mawson (MA) NM during a time period with changes in software, 2020 October through 2022 December, indicated by a colored trace for each counter tube.
(b) MA atmospheric pressure in hPa. 
 Vertical lines indicate the times of software changes. 
}
	\label{fig.MA_cali}
\end{figure}

\subsection{Mawson NM station}
Raw MA $L$ values are shown in Figure \ref{fig.MA_cali} from 2020 October through 2022 December.  Here we detail the primary modifications made to the data acquisition software. MA time delay histograms were first recorded using LandMonitor 8.161 starting in 2020 February, followed by a transition to LandMonitor 8.181 in 2021 June. Subsequent changes occurred with LandMonitor shifting to 8.182 in 2021 July and 8.183a in 2021 September. Around 2021 September--October, challenges emerged in writing histogram and pressure files, prompting a switch to LandMonitor 8.183n in 2021 November. Further updates transpired with LandMonitor changing to 8.183v, 8.183x, and 8.184 in 2021 December. 
Further changes ensued with LandMonitor adopting 8.185b and 8.185c in 2022 January. In 2022 June, LandMonitor transitioned to 8.186g. Presently, LandMonitor has been operating on version 8.186j, starting from 2022 September.

\end{document}